\documentclass[aps,prb,10pt,longbibliography,groupedaddress,twocolumn,letterpaper,nobalancelastpage,raggedbottom]{revtex4-2}

\usepackage{mycommands}

\begin{document}

    \newcommand{\articletitle}{Gyrotropy from Extrinsic Geometry in Twisted Materials}
    
    \title{\articletitle}

    \author{Spenser Talkington}
    \email{spenser@upenn.edu}
    \affiliation{Department of Physics and Astronomy, University of Pennsylvania, Philadelphia, Pennsylvania 19104, USA}
    
    \author{Eugene J. Mele}
    \affiliation{Department of Physics and Astronomy, University of Pennsylvania, Philadelphia, Pennsylvania 19104, USA}
    
    \date{\today}
    
    \begin{abstract}
        Gyrotropy in twisted bilayer graphene can be used as a signature of interlayer electronic coupling. Gyrotropy can emerge in the absence of interlayer coupling in time-reversal symmetric bilayer systems. This gyrotropy originates from the extrinsic geometry associated with the physical geometry of the system and is independent of the structure of the electronic states. We first illustrate this effect for a purely classical bilayer array of one-dimensional wires. Next we study pristine twisted bilayer graphene and show that the gyrotropy is entirely due to interlayer coupling. Finally we consider twisted bilayer MoTe$_2$, first as a pristine model where the gyrotropy exactly vanishes, and then with weak strain and displacement fields where we show that the geometric gyrotropy can dominate the gyrotropy due to interlayer coupling. Our results call attention to the necessity to separate the contribution of extrinsic physical geometry from the contribution of intrinsic electronic states to the properties of twisted materials.
    \end{abstract}
    
    \maketitle
    
    \section{Introduction}\label{sec:intro}

        Moir\'e bilayers with a relative interlayer twist have emerged as tunable platforms for realizing exotic physics associated with electronic coupling in a chiral structure. This intrinsic physics has emerged in platforms such as twisted bilayer graphene (TBG) near the magic angle \cite{morell2010flat,delaissardiere2010localization,bistritzer2011moire} where superconductivity \cite{cao2018unconventional,yankowitz2019tuning}, correlated insulators \cite{cao2018correlated} and other exotic behaviors are observed \cite{cao2020strange}, and in transition metal dichalcogenide (TMD) platforms \cite{wang2019magic,devakul2021magic,andrei2021marvels}. On the other hand, the twisted structure of these materials always introduces an ``extrinsic" geometric chirality from embedding in three-dimensional space, which has received relatively little attention. It is important to separate these contributions. In the limit of decoupled layers only the extrinsic effect remains, while with $C_{3z}$ and $\mathcal{T}$ symmetries only the intrinsic effect remains, allowing us to separate intrinsic and extrinsic contributions to the chiral optical gyrotropy.

        The optical activity of TBG has a long history with predictions for its response to linearly \cite{morell2012radiation,tabert2013optical,stauber2013optical,moon2013optical,stauber2023neutral} and circularly polarized light \cite{morell2012radiation,kim2016chiral,morell2017twisting,stauber2018chiral,stauber2018linear,addison2019twist,do2020optical,ho2023optical,qiu2024detection}, where the latter is sensitive to both the intrinsic electronic coupling and extrinsic geometric effects. In commensurate TBG \cite{shallcross2008quantum,mele2010commensuration,mele2012interlayer}, the competition between interlayer coupling effects and geometry should be particularly pronounced \cite{addison2019twist,talkington2023electric,talkington2023terahertz}. The chiral optical activity of TBG  has been measured in the far \cite{kim2016chiral} and near fields \cite{huang2022observation} and interpreted as an intrinsic effect due to interlayer coupling. Recently, the chiral optical activity of twisted WS$_2$ was measured in bilayers \cite{lan2021observation} and twisted stacks \cite{kim2024three}. To diagnose and interpret the origin of chiral optical activity we focus on the gyrotropy which is the linear in wavevector $q_z$, odd under inversion contribution to the transverse optical conductivity. In particular, the gyrotropy emerges due to a chiral optical conductivity $\sigma_{xy}^c$, a magnetoelectric coupling of electric and magnetic dipole currents as derived by Stauber \textit{et al} \cite{stauber2018chiral,stauber2018linear}. While naively this magnetoelectric coupling might be expected to be purely intrinsic and vanish in the absence of interlayer coupling we show that gyrotropy is generically present even with zero interlayer coupling due to extrinsic geometry.

        When interlayer coupling vanishes, a dipole current is conserved as identified by Nguyen and Son \cite{nguyen2020electrodynamics}, whose response to electric fields naturally decomposes into total, counterflow, and chiral optical conductivities \cite{stauber2018chiral,stauber2018linear,ochoa2020flat,ding2023chiral}. The counterflow current is perhaps most appreciated for its role in diagnosing flat bands \cite{bistritzer2010transport,bistritzer2011moire,stauber2023neutral,zhu2024layer}, but in our case it is a magnetic dipole current that responds to a total electric field in both layers. Further decompositions of the response are possible \cite{franta2020symmetry,pozo2023multipole} where the role of the embedding space on the electronic physics is particularly important \cite{avdoshkin2023extrinsic}.

        In this paper we first establish the general symmetry conditions for gyrotropy to be present in bilayers with particular consideration to moir\'e materials. In the presence of $C_{3z}$ and $\mathcal{T}$ the gyrotropy is purely due to interlayer coupling. Next, we present a minimal model of a classical bilayer array of isolated wires which breaks these symmetries and exhibits gyrotropy even in the absence of interlayer coupling. We then move on to TBG as the paradigmatic example of a twisted two-dimensional material and study its gyrotropy. Finally we consider twisted bilayer MoTe$_2$ as a representative example of a twisted TMD, and show conditions under which gyrotropy can emerge. We demonstrate that small heterostrain and a displacement field allow gyrotropy and it is almost entirely extrinsic/independent of interlayer coupling. Overall we identify separate contributions to gyrotropy from intrinsic electronic coupling and extrinsic geometric effects with the key conclusion that interlayer electronic coupling is generically not responsible for the full response properties and real-space geometry makes a significant contribution.

    \section{(Anti)Symmetric Bilayer Currents}

        In a bilayer system with bottom layer $B$ and top layer $T$ we generically have the relation between layer-resolved currents $\bm{j}^l=(j_x^l,j_y^l)$ and electric fields $\bm{E}^l=(E_x^l,E_y^l)$
        \begin{align}
            \begin{pmatrix} \bm{j}^B\\\bm{j}^T \end{pmatrix}
            = \begin{pmatrix}
                \sigma^{BB} & \sigma^{BT}\\
                \sigma^{TB} & \sigma^{TT}
            \end{pmatrix}
            \begin{pmatrix} \bm{E}^B\\\bm{E}^T \end{pmatrix},
        \end{align}
        which defines the conductivity matrix $\sigma$ via $\bm{j}=\sigma \bm{E}$.
        
        Let us separate the layer symmetric ``total" electric dipole type current $\bm{j}^{tot}=\bm{j}^T+\bm{j}^B$ and layer antisymmetric ``counterflow" magnetic dipole type current $\bm{j}^{cf}=\bm{j}^T-\bm{j}^B$, and the associated electric fields $\bm{E}^{tot}=(\bm{E}^T+\bm{E}^B)/2$ and $\bm{E}^{cf}=(\bm{E}^T-\bm{E}^B)/2$, with the corresponding relation between current and electric field
        \begin{align}
            \begin{pmatrix} \bm{j}^{tot}\\\bm{j}^{cf} \end{pmatrix}
            = \begin{pmatrix}
                \sigma^{tot} & \sigma^{c}\\
                \sigma^{c'} & \sigma^{cf}
            \end{pmatrix}
            \begin{pmatrix} \bm{E}^{tot}\\\bm{E}^{cf} \end{pmatrix}.
        \end{align}
        After some straightforward algebra one finds $\Sigma=\mathcal{H}\sigma \mathcal{H}^\top$ with Hadamard-like matrix $\mathcal{H}$
        \begin{align}\label{eq:hadamard}
            \begin{pmatrix}
                \sigma^{tot} & \sigma^{c}\\
                \sigma^{c'} & \sigma^{cf}
            \end{pmatrix} = \begin{pmatrix}
                \mathbbm{1}&\mathbbm{1}\\-\mathbbm{1}&\mathbbm{1}
            \end{pmatrix} \begin{pmatrix}
                \sigma^{BB} & \sigma^{BT}\\
                \sigma^{TB} & \sigma^{TT}
            \end{pmatrix} \begin{pmatrix}
                \mathbbm{1}&\mathbbm{1}\\-\mathbbm{1}&\mathbbm{1}
            \end{pmatrix}^\top
        \end{align}
        
        From this we see the blocks $\sigma^{tot}$ and $\sigma^{cf}$ where layer aligned/opposite fields induce layer aligned/opposed currents. Our focus will be on the ``chiral" components $\sigma^c$ where a counterflow electric field can induce a total in-plane current. This can be rewritten as an effective electric-magnetic dipole magnetoelectric coupling: opposite in-plane electric fields induce an in-plane magnetic moment, and an in-plane magnetic field induces a total in-plane current \cite{stauber2018chiral,wang2020optical}. This can also be understood as a spatially dispersive response dependent on wavevector $q_z$ and interlayer separation $d_z$ as $d_z\partial_z \bm{E}\sim i q_z d_z\, \bm{E}^{tot}$ \cite{nguyen2020electrodynamics,mahon2020magnetoelectric,pozo2023multipole}. The component $\sigma_{xy}^c$ is a gyrotropic natural optical activity of a thin film; $\sigma_{xy}^{tot}$ enters at order $q_z^0$ and is an ordinary transverse response, while the optical activity due to $\sigma_{xy}^{cf}$ enters at order $q_z^2$ and is a sub-leading response. The real/dissipative part of $i q_z d_z\, \sigma_{xy}^c$ causes circular dichroism and ellipticity, while its imaginary/reactive part causes circular birefringence and polarization rotation. Hereafter we call $\sigma_{xy}^c$ the gyrotropy.

        \subsection{Intrinsic Gyrotropy in Twisted Bilayers}

        As an example, let us analyze some of the symmetry properties of the conductivity tensor with an eye towards twisted materials. Particularly relevant symmetries are time-reversal symmetry $\mathcal{T}$ and in-plane $2\pi/3$ rotation $C_{3z}$. Twisted bilayer graphene and twisted bilayer transition metal dichalcogenides (TMDs) like MoTe$_2$, WS$_2$ and WSe$_2$ have these symmetries in the absence of explicit or spontaneous symmetry breaking \cite{zou2018band}. Some other moir\'e materials such as 1T$'$-WTe$_2$ \cite{yu2024strongly,kawakami2026one}, CrSBr \cite{liu2025moire,li2025magneto}, GeSe \cite{kennes2020one}, and phosphorene \cite{zhao2021anisotropic,soltero2022moire,jiang2026twist} break $C_{3z}$ \cite{wang2023tunable} and would have a richer response structure.

        Let us consider $\sigma_{xx}^c$ and $\sigma_{yy}^c$. First, $C_{3z}$ symmetry imposes full rotational symmetry at $q=0$ so that $\sigma_{xx}^{ll'}=\sigma_{yy}^{ll'}$ and hence $\sigma_{xx}^c=\sigma_{yy}^c$. Next Onsager $\mathcal{T}$ reciprocity gives $\sigma_{xx}^{TB}=\sigma_{xx}^{BT}$. Therefore with $C_{3z}$ and $\mathcal{T}$
        \begin{align}\label{eq:sxxc-c3zT}
            \sigma_{xx}^c = \sigma_{xx}^{TT}-\sigma_{xx}^{BB}.
        \end{align}
        Moreover there is often a layer exchange symmetry, $C_{2x}$ imposes that $\sigma_{xx}^{TT}=\sigma_{xx}^{BB}$, in which case $\sigma_{xx}^c=0$.

        Next let us consider $\sigma_{xy}^c$. The valley-resolved conductivity $\sigma^{ll'}_{xy}(K)$ can be decomposed into Hall and symmetric parts $\sigma^{ll'}_{xy}(K)=\sigma^{ll'}_H(K)+\sigma^{ll'}_S(K)$. Generically $C_{3z}$ imposes that $\sigma_{yx}^{ll'}(K)=-\sigma_{xy}^{ll'}(K)$ and $\sigma_{xy}^{ll'}(K)=\sigma_H^{ll'}(K)$ with vanishing $\sigma_S^{ll'}(K)$. Time reversal maps $K$ onto $K'$ and the valley-resolved Onsager $\mathcal{T}$ relation \cite{casimir1945onsager} imposes that $\sigma_{\alpha\beta}^{ll'}(K)=\sigma_{\beta\alpha}^{l'l}(K')$ so with the antisymmetry of the Hall response this gives $\sigma_{xy}^{ll'}(K)=-\sigma_{xy}^{l'l}(K')$. Using this, we immediately see that summing over valleys gives $\sigma_{xy}^{TT}=\sigma_{xy}^{BB}=0$. Moreover, it tells us that $\sigma_{xy}^{TB}=\sigma_{xy}^{TB}(K)+\sigma_{xy}^{TB}(K')=-\sigma_{xy}^{BT}(K')-\sigma_{xy}^{BT}(K)=-\sigma_{xy}^{BT}$. Therefore for systems with $C_{3z}$ and $\mathcal{T}$
        \begin{align}\label{eq:sxyc-c3zT}
            \sigma_{xy}^c = 2\,\sigma_{xy}^{BT}.
        \end{align}
        This tells us that in the presence of $C_{3z}$ and $\mathcal{T}$ for zero interlayer coupling both $\sigma^c_{xy}$ and $\sigma^{c'}_{xy}$ vanish. So any chiral optical response $\sigma_{xy}^c$ in twisted bilayer graphene or MoTe$_2$ with these symmetries is an \textit{intrinsic} gyrotropy due to electronic coupling through interlayer hybridization or result from symmetry breaking.
        
        \begin{figure}
            \centering
            \includegraphics[width=\linewidth]{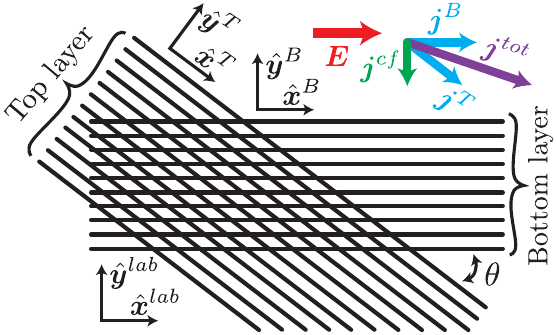}
            \caption{A bilayer array of classical wires with relative twist $\theta$ between layers and no interlayer coupling exhibits finite gyrotropy, $\sigma_{xy}^{c'}=\sigma_{xy}^c$ in this model. This gyrotropy means that a uniform electric field can drive a transverse (magnetic dipole) current: $\bm{E}$ (red) is uniform in both layers, and so the antisymmetric electric field $\bm{E}^{cf}$ is zero, but $\bm{j}^{cf}=\bm{j}^T-\bm{j}^B$ (green) is still finite and transverse to $\bm{E}$. This ``geometric gyrotropy" depends solely on the \textit{extrinsic} physical geometry and not on the \textit{intrinsic} interlayer electronic coupling.}
            \label{fig:wires}
        \end{figure}
        
    \section{Bilayer Arrays of 1D Wires}

        As an example, let us consider building a 2D material from an array of 1D wires with spacing $\ell$ and conductivity $\sigma_{xx}^{1D}$ (e.g. Drude conductivity) where the $xx$ part of the conductivity is $\sigma_{xx}^{2D}=\sigma_{xx}^{1D}/\ell$ and all other components vanish. Note that this model explicitly breaks $C_{3z}$ and therefore supports geometric gyrotropy, similar to the rotors studied in Refs. \cite{rogacheva2006giant,plum2007giant}. We illustrate our setup in Fig. \ref{fig:wires}. If we introduce a second layer with a relative twist $\theta$, the conductivity of the ``top" layer is
        \begin{align}
            \sigma^{TT} = \sigma_{xx}^{2D} \begin{pmatrix} \cos^2(\theta) & \cos(\theta)\sin(\theta)\\
            \cos(\theta)\sin(\theta) & \sin^2(\theta) \end{pmatrix},
        \end{align}
        or at small twist angles to linear order in $\theta$
        \begin{align}
            \sigma^{TT} \approx \sigma_{xx}^{2D} \begin{pmatrix} 1 & \theta\\
            \theta & 0 \end{pmatrix}.
        \end{align}
        The ``bottom" layer remains untwisted and has $\sigma^{BB}=\sigma^{2D}$. For no interlayer coupling, $\sigma^{TB}=\sigma^{BT}=0$, so $\sigma^{tot}=\sigma^{cf}$ and $\sigma^c=\sigma^{c'}$ with
        \begin{align}
            \sigma^{tot} \approx \sigma_{xx}^{2D} \begin{pmatrix}
                2&\theta\\\theta&0
            \end{pmatrix}
            , \quad
            \sigma^c \approx \sigma_{xx}^{2D} \begin{pmatrix}
                0&\theta\\\theta&0
            \end{pmatrix},
        \end{align}
        where we see that even in the absence of interlayer coupling twisted layers have a linear in $\theta$ chiral transverse component, $\sigma_{xy}^c$. This contribution is entirely \textit{extrinsic} to the geometry of the material and not \textit{intrinsic} to the electronic states in this limit; we label this contribution ``geometric gyrotropy". We illustrate this effect using the arrows in Fig. \ref{fig:wires}: non-zero $\sigma_{xy}^{c'}$ means that a uniform electric field (red) drives a transverse counterflow current (green).

    \section{Twisted Bilayer Graphene}

        Next, we consider the case of a truly two-dimensional material with twisted bilayer graphene (TBG). TBG is the prototypical twisted two-dimensional material where two honeycomb lattices of carbon atoms are stacked with a relative interlayer twist $\theta$. TBG can be treated atomistically using tight-binding models \cite{morell2010flat,delaissardiere2010localization}, but a continuum model where the matrix size is independent of twist angle is preferable. Such a continuum model is given in \cite{dossantos2007graphene,bistritzer2010transport} and its extension in \cite{dossantos2012continuum}. We will be interested in small twist angles but away from the flat band limit where interactions do not need to be taken into account. In this strong layer coupling case one needs to go beyond the first shell of momentum scatterings considered by Lopes dos Santos et al in \cite{dossantos2007graphene}, and use the momentum-space lattice of scatterings considered by Bistritzer and MacDonald \cite{bistritzer2011moire}. We thus refer to this continuum model for TBG with truncation beyond the first shell as the Bistritzer-MacDonald continuum model.

        \subsection{Bistritzer-MacDonald Continuum Model}

            Here we present the Bistritzer-MacDonald (BM) continuum model for TBG \cite{dossantos2007graphene,bistritzer2011moire}. This model accurately captures the low-energy physics of TBG including its magic angle at $\sim 1.1^\circ$ interlayer twist. In a recent study \cite{xiao2026imaging}, Xiao, \textit{et al} showed that the band structure of TBG is correctly captured by the non-interacting BM model even at $1.2^\circ$ using the quantum twisting microscope \cite{inbar2023quantum,wei2025dirac}. The quantum twisting microscope can also resolve electron-phonon interactions \cite{birkbeck2025quantum} and renormalization of the Fermi velocity by interactions in the monolayer \cite{lee2026revealing}. For definiteness we consider $2^\circ$ which is both a small angle and should be outside the interaction-dominated regime.

            The $K$-valley BM model is
            \begin{align} \label{eq:H_full}
                H_K =
                \begin{pmatrix}
                    H^{B} + V^{B} & H^{BT}\\
                    H^{TB} & H^{T} + V^{T}
                \end{pmatrix},
            \end{align}
            with size $4(2N+1)^2$ when scatterings with up to $N$ reciprocal lattice vectors are included, where the $N$ required for convergence depends on twist angle. We fix $N=2$ since as we have verified numerically this produces the same results for the gyrotropy as $N=3$ and $4$ to within 1\%. The $K'$-valley model is related to the $K$-valley model by $\mathcal{T}$ so $H_{K'}(\bm{k})=H_K^*(-\bm{k})$. The intralayer terms are
            \begin{align}\label{eq:H_intra}
                H^{l} = \!\!\!\!\sum_{n,m=-N,\dots, N}\!\!\!\! h_{nm}^{l}(\bm{k}) c^\dag_{l,\bm{k}+n\bm{g}_1+m\bm{g}_2} c_{l,\bm{k}+n\bm{g}_1+m\bm{g}_2},
            \end{align}
            where the Hamiltonian is a $2\times 2$ sublattice continuum Hamiltonian describing linearly dispersing Dirac cones $h_{nm}^{l}=t_\parallel \bm{\sigma}^{l}\cdot(\bm{k}-\bm{\kappa}^{l}+n\bm{g}_1+m\bm{g}_2)$ with $t_\parallel=0.52~\mathrm{eV\cdot nm}$ and layer Pauli matrix vector $\bm{\sigma}^l=(\sigma_x^l,\sigma_y^l)$.
            We have introduced the reciprocal lattice vectors $\bm{g}_1=|\bm{K}_{mono}|\sin(\theta/2)(\sqrt{3},3)$, $\bm{g}_2=|\bm{K}_{mono}|\sin(\theta/2)(-\sqrt{3},3)$ with rotated Dirac point positions $\bm{\kappa}_\pm = \pm |\bm{K}_{mono}|\sin(\theta/2)(0,1)$, $\bm{\kappa}^{B}=\bm{\kappa}_+$ and $\bm{\kappa}^{T}=\bm{\kappa}_-$, where $\bm{K}_{mono}=4\pi(1,0)/3a$ for graphene's lattice constant $a=0.246$ nm.
            All atoms are the same, and so the layer-dependent potential $V^{l}$ is small can be set to zero.
            The layer Pauli matrices are
            \begin{align}
                \sigma_i^B = \rho_B\sigma_i\rho_B^\dag,\quad \sigma_i^T = \rho_T\sigma_i\rho_T^\dag,
            \end{align}
            for $i=x,y$ with $\rho_B=e^{i\theta\sigma_z/4}$, $\rho_T=e^{-i\theta\sigma_z/4}$ corresponding to rotating the bottom layer $\theta/2$ and top layer $-\theta/2$.
        
            To complete the BM model the interlayer tunneling terms are
            \begin{align}\label{eq:H_inter}
                H^{TB} = \!\!\!\!\sum_{n,m=-N,\dots,N}\!\!\!\! &T_0\, c^\dag_{T,\bm{k}+n\bm{g}_1+m\bm{g}_2} c_{B,\bm{k}+n\bm{g}_1+m\bm{g}_2}\\
                +&T_1\, c^\dag_{T,\bm{k}+(n-1)\bm{g}_1+m\bm{g}_2} c_{B,\bm{k}+n\bm{g}_1+m\bm{g}_2}\notag\\
                +&T_2\, c^\dag_{T,\bm{k}+n\bm{g}_1+(m-1)\bm{g}_2} c_{B,\bm{k}+n\bm{g}_1+m\bm{g}_2},\notag
            \end{align}
            where $H^{BT}=(H^{TB})^\dag$, and terms whose shifted indices fall outside the cutoff window are omitted.
            The $T_i$ are matrices in sublattice space given by
            \begin{align}
                T_0 \!=\! \begin{pmatrix} t_0 & \!\!t_1\\ t_1 & \!\!t_0 \end{pmatrix}\!,\
                T_1 \!=\! \begin{pmatrix} t_0 \zeta^* & \!\!t_1\\ t_1 \zeta & \!\!t_0 \zeta^* \end{pmatrix}\!,\ T_2 \!=\! \begin{pmatrix} t_0 \zeta & \!\!t_1 \\ t_1 \zeta^* & \!\!t_0 \zeta \end{pmatrix},
            \end{align}
            where $\zeta=e^{2\pi i/3}$ and $t_0$ quantifies tunneling from one sublattice in one layer to the same sublattice in the other layer, and $t_1$ quantifies tunneling between different sublattices. We will take $t_0=t_1=t_\perp$ which corresponds to considering encapsulated samples with suppressed out of plane breathing \cite{nam2017lattice,ezzi2024analytical}; for the physical case $t_\perp\approx 0.11$ eV.

        \subsection{Optical Conductivity}
        
            We use the layer-resolved Kubo formula \cite{mahan2000many}
            \begin{align}\label{eq:kubo}
                \sigma_{\alpha\beta}^{ll'} \!&=\! i\hbar\! \int_{BZ}\! \frac{d^2k}{(2\pi)^2} \sum_{s}\! \bigg[ -\frac{\partial f(E_s)}{\partial E_s} \frac{\langle u_s|j_\alpha^l|u_{s}\rangle\langle u_{s}|j_\beta^{l'}|u_s\rangle}{\hbar\omega+i\eta}\notag \\
                &\!+\sum_{s'\neq s}\! \frac{(f(E_s)\!-\!f(E_{s'}))}{E_{s'}-E_s}\frac{\langle u_s|j_\alpha^l|u_{s'}\rangle\langle u_{s'}|j_\beta^{l'}|u_s\rangle}{\hbar\omega-(E_{s'}-E_s)+i\eta}\bigg],
            \end{align}
            where the first line gives the intraband contribution and the second line gives the interband contribution. $\hbar\omega$ is the photon energy, $\eta$ is a phenomenological broadening, $f(E)=1/(e^{(E-\mu)/k_B T}\!+\!1)$, and current operators are $j_\alpha^l = e\, P_l (\partial H/\hbar\, \partial k_\alpha) P_l$ for projector $P_l$ onto layer $l$ where $e$ is the electron charge. The single-particle states satisfy $H|u_s\rangle = E_s |u_s\rangle$ for $H=H_K$ or $H_{K'}$ where we obtain the physical optical conductivity by summing the conductivities for valleys $K$ and $K'$, and multiplying by the spin degeneracy $g_s=2$. For the TMD model described below, spin-valley locking means $g_s=1$. To numerically integrate, we sum over momenta on a uniform $60\times 60$ mesh in the Brillouin zone for all conductivity calculations; we have confirmed that this mesh is converged. For the BM model we take $T=77~\mathrm{K}$ and $\eta=0.020$ eV, while for the TMD model we take $T=10~\mathrm{K}$ and $\eta=5$ meV, which are chosen so that the responses are smooth over the frequency range plotted.

        \begin{figure}
            \centering
            \includegraphics[width=\linewidth]{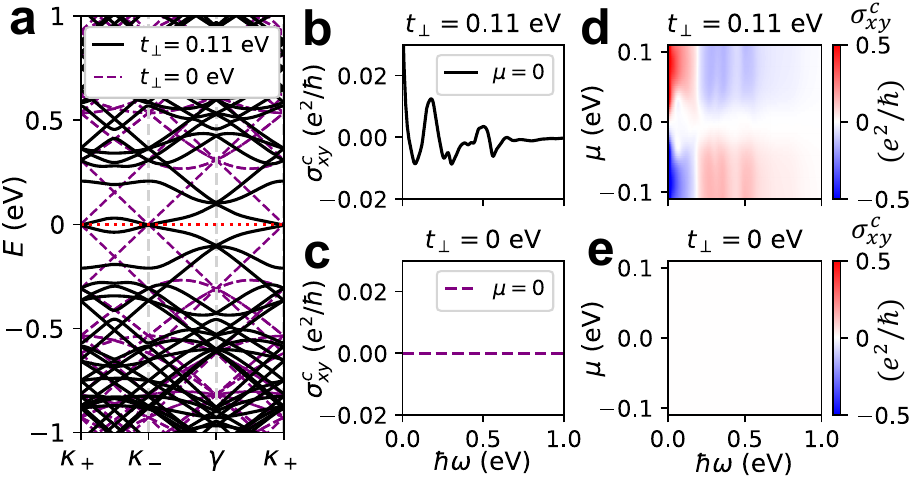}
            \caption{Chiral optical conductivity of twisted bilayer graphene (TBG) at $2^\circ$ as a function of interlayer coupling $t_\perp$. For real TBG, $t_\perp\approx0.11$ eV. \textbf{(a)} Band structure: $t_\perp=0.11$ eV is in black, $t_\perp=0$ is dashed in purple; note the emergence of the Van Hove singularity between $\bm{\kappa}_+$ and $\bm{\kappa}_-$ at finite $t_\perp$. Chemical potential is at charge neutrality (dotted red line). \textbf{(b-c)} Gyrotropies $\sigma_{xy}^c$ for $\mu=0$ with $t_\perp=0.11$ eV and 0 respectively. \textbf{(d-e)} Gyrotropy $\sigma_{xy}^c$ as a function of frequency and chemical potential for (d) physical and (e) zero interlayer coupling. The response is zero for zero interlayer coupling as enforced by Eq. \ref{eq:sxyc-c3zT}. The crossover from positive to negative in (d) as a function of frequency originates with the competition of intraband and interband contributions entering with opposite signs.}
            \label{fig:tbg}
        \end{figure}
        
        \subsection{Results}

            In Fig. \ref{fig:tbg}(b-c) we see that at finite interlayer coupling the gyrotropy is dominated by inter Van Hove singularity transitions at finite frequency \cite{kim2016chiral}. At zero interlayer coupling the gyrotropy vanishes as required by the presence of $C_{3z}$ and $\mathcal{T}$. While a truncation to twist-angle independent $H$ and current operators would be internally consistent, it would completely miss the gyrotropy as it enters at linear order in twist angle. The continuum model as written has particle-hole symmetry, a point-like Fermi surface, and the response is weak at charge neutrality. To enhance the response we can consider finite doping to enlarge the Fermi surface, Fig. \ref{fig:tbg}(d-e) where we see at finite interlayer coupling and doping, $\sigma_{xy}^c$ can be of order $e^2/\hbar$, while at zero interlayer coupling the response remains zero as imposed by Eq. \ref{eq:sxyc-c3zT}.
    
    \section{Twisted Bilayer \texorpdfstring{$\mathbf{MoTe_2}$}{plain}}

        Next, we consider the case of a $K$-valley moir\'e model of twisted bilayer transition metal dichalcogenide (TMD), with a single active sublattice. As with TBG, this can be treated atomistically using tight-binding models \cite{wu2018hubbard,jia2024moire}, but at small twist angles this becomes infeasible and continuum models where the matrix size is independent of twist angle are preferable. Wu, Lovorn, Tutuc, Martin and MacDonald constructed such a continuum model \cite{wu2019topological}. These TMDs have one transition metal atom on one sublattice of the honeycomb lattice and two chalcogenide atoms on the other sublattice. This breaks the sublattice symmetry and simplifies the model to a single scalar per valley \cite{xu2014spin}. We will consider MoTe$_2$, although we could equally well have focused on WSe$_2$. Twisted bilayer MoTe$_2$ has been studied with great interest recently due to its strongly interacting electronic physics at small twist angles: ferromagnetism and spontaneous time-reversal symmetry breaking are observed from 2.1 to 4.5$^\circ$ \cite{li2025universal,zeng2023integer,kang2024evidence,wu2026observation}, with fractional Chern insulators seen between 3.5 and 3.9$^\circ$ \cite{cai2023signatures,park2023observation}, and other exotic physics is seen at these twist angles. Here we choose 5$^\circ$ to avoid the strong correlation and spontaneous $\mathcal{T}$-breaking regime so that we can consider a non-interacting model, but the extension of our approach to spontaneous time-reversal symmetry breaking is an interesting question. We first introduce the continuum model, identify that the gyrotropy is enforced to vanish by symmetry, and then enable gyrotropy by applying heterostrain and a displacement field.

        \subsection{Continuum Model}

            The $K$-valley TMD continuum model is the same as in Eq. \ref{eq:H_full} but with size $2(2N+1)^2$ since the model only has one sublattice. As in the BM model, the $K'$-valley model is related to the $K$-valley model by $H_{K'}(\bm{k})=H_K^*(-\bm{k})$. The intralayer terms have the form of Eq. \ref{eq:H_intra} where the Hamiltonian is the quadratically dispersing valence band below the semiconducting gap, $h_{nm}^{l}=-(\hbar^2/2m^*)\,|\bm{k}-\bm{\kappa}^{l}+n\bm{g}_1+m\bm{g}_2|^2$ with $\hbar^2/2m^*=61.45~\mathrm{meV\cdot nm^2}$ for MoTe$_2$ ($m^*=0.62\, m_e$) and where the lattice constant is $a=0.3472$ nm and $m_e$ is the free electron mass. Here $V^{l}$ is a layer-dependent potential that accounts for the spatial variation of the local interlayer stacking which is irrelevant in TBG, but important in TMDs. Specifically $V=8$ meV and $\psi=-89.6^\circ$ for MoTe$_2$ \cite{wu2019topological} where
            \begin{align}
                V^{l} = \!\!\!\!\!\!\! \sum_{n,m=-N,\dots, N}\!\!\!\!\!\!\! Ve^{ is_l\psi}\big[&c^\dag_{l,\bm{k}+n\bm{g}_1+m\bm{g}_2} c_{l,\bm{k}+n\bm{g}_1+(m+1)\bm{g}_2} \notag\\
                +&c^\dag_{l,\bm{k}+n\bm{g}_1+m\bm{g}_2} c_{l,\bm{k}+(n-1)\bm{g}_1+m\bm{g}_2} \notag\\
                +&c^\dag_{l,\bm{k}+n\bm{g}_1+m\bm{g}_2} c_{l,\bm{k}+(n+1)\bm{g}_1+(m-1)\bm{g}_2} \big]\notag\\& \hspace{-0.5 in} + h.c. \,,
            \end{align}
            which sums over the first star of $\bm{g}$ vectors and $s_B=1$, $s_T=-1$. As in the BM model, we fix $N=2$. Terms whose shifted indices fall outside the cutoff window are omitted. To complete the TMD continuum model, the interlayer tunneling terms are those in Eq. \ref{eq:H_inter} where $T_i$ are scalars due to the effective one-sublattice model. For MoTe$_2$ we choose $T_0=T_1=T_2=t_\perp= -8.5$ meV which quantifies tunneling between layers and accounts for out-of-plane lattice relaxation, but ignores in-plane relaxation which is dependent on twist angle \cite{wu2019topological}.

            For the current operators, as in twisted bilayer graphene, we take $j_\alpha^l=e\, P_l(\partial H/\hbar\,\partial k_\alpha)P_l$ which evaluate to
            \[
            \bm{j}^l
            = -\frac{e\hbar}{m^*} (\bm{k}-\bm{\kappa}^l+n\bm{g}_1+m\bm{g}_2)
            \]
            where $\bm{\kappa}^l$ and $\bm{g}_i$ and the wavefunctions in Eq. \ref{eq:kubo} all include the effects of twist and heterostrain.

        \begin{figure}
            \centering
            \includegraphics[width=\linewidth]{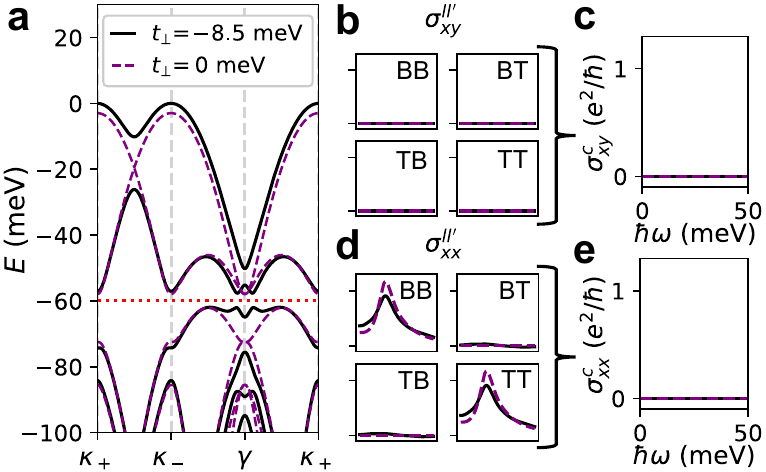}
            \caption{Twisted bilayer MoTe$_2$ lacks chiral optical activity in the continuum model. \textbf{(a)} Band structure at $5^\circ$ for interlayer tunneling, $t_\perp$, equal to its physical value (black), and to zero (purple); the chemical potential $\mu=-60$ meV is indicated by the dotted red line. \textbf{(b-c)} Layer-resolved and chiral $\sigma_{xy}$ both exactly vanish due to the combination of $C_{3z}$, $\mathcal{T}$ and accidental antiunitary layer-exchange symmetry $\Lambda$ as detailed in the main text. \textbf{(d-e)} Layer-resolved $\sigma_{xx}$ is non-vanishing, but $C_{2x}$ constrains $\sigma_{xx}^{TT}=\sigma_{xx}^{BB}$ and Onsager $\mathcal{T}$ reciprocity ensures $\sigma_{xx}^{TB}=\sigma_{xx}^{BT}$; therefore $\sigma_{xx}^c$ vanishes.}
            \label{fig:tmd}
        \end{figure}

        \subsection{Vanishing Response}

            A priori, one might expect that the twisted TMD model would also exhibit finite chiral optical conductivity due to its twisted structure. We show that this is not the case, even in the presence of interlayer coupling. The reason is that three-fold rotation symmetry with a scalar sublattice structure is not anisotropic enough to allow a gyrotropy. From Eq. \ref{eq:sxxc-c3zT} it is evident that $\sigma_{xx}^c=0$ since the pristine continuum model has $C_{3z}$, $\mathcal{T}$, and $C_{2x}$ symmetries. What is less obvious is why $\sigma_{xy}^c$ vanishes. We begin by recalling Eq. \ref{eq:sxyc-c3zT}: $\sigma_{xy}^c = 2\sigma_{xy}^{BT}$ and then show $\sigma_{xy}^{BT}=0$. First, we note that the TMD model has an accidental scalar antiunitary layer-exchange symmetry that is absent in the BM model given by
            \begin{align}
                \lambda H_K^*(\bm{k}) \lambda^{-1} = H_K(-\bm{k}),
            \end{align}
            for layer exchange $\lambda$ taking $T\leftrightarrow B$. This symmetry corresponds to an operator $\Lambda = \lambda \widetilde{\mathcal{K}}$ where $\widetilde{\mathcal{K}}$ acts as complex conjugation on scalars and reverses momenta, but does not relabel $\bm{G}$ vectors; on the standard layer, moir\'e $\bm{G}$-vector, momentum-resolved plane-wave states $\Lambda$ acts as $\Lambda |l,\bm{G},\bm{k}\rangle = |\lambda l,\bm{G},-\bm{k}\rangle$. This antiunitary symmetry has a corresponding reciprocity relation $\sigma_{\alpha\beta}^{ll'}(K)=\sigma_{\beta\alpha}^{(\lambda l')(\lambda l)}(K)=\sigma_{\beta\alpha}^{ll'}(K)$ for $l\neq l'$ so $\sigma_{xy}^{BT}(K)=\sigma_{yx}^{BT}(K)$. Next, due to $C_{3z}$ symmetry $\sigma_{xy}^{BT}(K)=\sigma_H^{BT}(K)=-\sigma_{yx}^{BT}(K)$. Combining $\Lambda$ and $C_{3z}$, $\sigma_{yx}^{BT}(K)=-\sigma_{yx}^{BT}(K)$ so $\sigma_{yx}^{BT}(K)=0$. Thus $\sigma_{xy}^{BT}(K)=0$ and $\mathcal{T}$ ensures that a similar argument holds for $K'$ so $\sigma_{xy}^c=0$ for the pristine TMD model.

            We see these symmetry constraints on the conductivity are borne out in Fig. \ref{fig:tmd}. In Fig. \ref{fig:tmd}(b-c) we see that all layer-resolved $\sigma_{xy}^{ll'}$ vanish and so $\sigma_{xy}^c=0$. In Fig. \ref{fig:tmd}(d-e) we see that $\sigma_{xx}^{TT}=\sigma_{xx}^{BB}$ and $\sigma_{xx}^{TB}=\sigma_{xx}^{BT}$ as expected and so $\sigma_{xx}^c$ vanishes.  While this model is pristine, realistic devices have finite heterostrain and applying a displacement field is standard; incorporating these two elements breaks both $C_{3z}$ and $\Lambda$ and allows a gyrotropy in the absence of $\mathcal{T}$ breaking as we will see below.

        \subsection{Heterostrain}

            Here we will consider heterostrain which acts to rigidly distort the lattice. Although homostrain breaks $C_{3z}$, it acts identically on the two layers and cannot generate $\sigma_{xy}^c$; therefore, we need to consider heterostrain. These effects naturally emerge in real devices and their effects on the optical conductivity of TBG were considered in Ref. \cite{dai2021effects}. We begin with the rotation matrix for rotation $\theta$ about the $z$-axis
            \begin{align}
                R(\theta) =
                \begin{pmatrix}
                    \cos(\theta) & -\sin(\theta)\\
                    \sin(\theta) & \cos(\theta)
                \end{pmatrix},
            \end{align}
            and the strain matrix for strain $\epsilon$ oriented at angle $\theta_s$ is \cite{bi2019designing,escudero2024designing}
            \begin{align}
                S(\epsilon,\theta_s) = \mathbbm{1} + \epsilon\, \bm{d} \bm{d}^\top \!\!- \nu \epsilon\, \bm{d}_\perp \bm{d}_\perp^\top,
            \end{align}
            in terms of the direction $\bm{d}=(\cos(\theta_s),\sin(\theta_s))$, orthogonal direction $\bm{d}_\perp=(-\sin(\theta_s),\cos(\theta_s))$ and Poisson ratio $\nu=0.25$ for MoTe$_2$ \cite{mortazavi2018mechanical,woo2016poisson}. We consider the deformation of rotating the top layer by $-\theta/2$ and the bottom layer by $\theta/2$, and then straining the top layer by $-\epsilon/2$ and the bottom layer by $\epsilon/2$ along direction $\bm{d}$. This corresponds to the transformation
            \begin{align}
                \mathscr{T}_B \!=\! S(\epsilon/2, \theta_s) R(\theta/2)
                ,\,
                \mathscr{T}_T \!=\! S(-\epsilon/2, \theta_s) R(-\theta/2).
            \end{align}
            Real-space vectors transform as $\bm{a}_i\to \mathscr{T} \bm{a}_i$, while reciprocal-space vectors transform as $\bm{b}_i\to (\mathscr{T}^{-1})^\top \bm{b}_i$.

            We consider $\bm{K}_{mono}=4\pi(1,0)/3a$ as before and then we have the monolayer reciprocal lattice vectors $\bm{b}_\pm = |\bm{K}_{mono}|(3,\pm\sqrt{3})/2$ from which $\bm{b}_\pm^{B/T}=(\mathscr{T}_{B/T}^{-1})^\top \bm{b}_\pm$. Thence the moir\'e reciprocal lattice vectors are $\bm{g}_1 = \bm{b}_-^B-\bm{b}_-^T$ and $\bm{g}_2 = \bm{b}_+^B-\bm{b}_+^T$. Likewise $\bm{K}^{B/T}=(\mathscr{T}_{B/T}^{-1})^\top \bm{K}_{mono}$ from which $\bm{\kappa}_\pm=\pm (\bm{K}^{B}-\bm{K}^{T})/2$.

        \begin{figure}
            \centering
            \includegraphics[width=\linewidth]{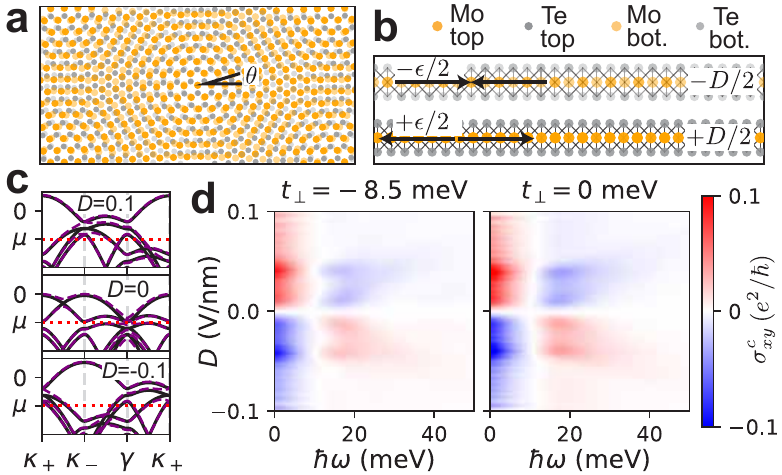}
            \caption{Chiral optical activity can appear in twisted bilayer MoTe$_2$ under weak heterostrain $\epsilon$ in the presence of a displacement field $D$. Heterostrain breaks $C_{3z}$, but this is insufficient to enable optical gyrotropy as $\lambda \tau_z \lambda^{-1} = -\tau_z$ imposes that the gyrotropy is odd in $D$ and so at $D=0$ the response is zero. \textbf{(a)} Top view of bilayer MoTe$_2$ with a $5^\circ$ relative twist and $0.1\%$ heterostrain. \textbf{(b)} Side view of zigzag direction illustrating strain direction and displacement field. \textbf{(c)} Band structure for select displacement fields. Chemical potential $\mu=-60$ meV is indicated in red, and as before black lines are for the physical $t_\perp$, while dashed purple lines are for the decoupled bilayer. \textbf{(d)} Chiral optical activity for both the full response at the physical $t_\perp$ and the decoupled layer response. The decoupled response---extrinsic gyrotropy---is the primary component of the full physical response and it dominates the gyrotropy intrinsic to interlayer coupling.}
            \label{fig:tmd-strain}
        \end{figure}    

        \subsection{Response with Heterostrain and Displacement Field}

            Applying heterostrain breaks $C_{3z}$ and applying a displacement field breaks layer exchange symmetry and hence $\Lambda$ so gyrotropy is symmetry allowed. We consider the case of weak heterostrain along the $x$-axis with $0.1\%$ strain between the two layers; this value is typical of tear-and-stack devices which routinely exhibit heterostrains of up to 1\% \cite{thompson2025microscopic}. We illustrate the setup in Fig. \ref{fig:tmd-strain}(a-b) where the layers are both rotated relative to each other and strained. We consider a displacement field, $(ed_zD/2)\tau_z$ with Pauli-$z$ matrix $\tau_z$ acting in layer space, electron charge $e$ and interlayer spacing $d_z=0.69$ nm. We see in Fig. \ref{fig:tmd-strain}(c) that the bands shift and the band maximum shifts from two equal peaks at $\bm{\kappa}_+$ and $\bm{\kappa}_-$ to moir\'e-valley anisotropy with a maximum at $\bm{\kappa}_+$ for positive $D$ and a maximum at $\bm{\kappa}_-$ for negative $D$.

            With these two ingredients we see a response in Fig. \ref{fig:tmd-strain}(d). We note several features of the response. First, $\sigma_{xy}^c$ is odd in $D$, corresponding to the reversal of layer polarization under $D\to -D$ and the fact that $\sigma_{xy}^c$ is odd under $T\leftrightarrow B$; this follows from $\Lambda H(\bm{k},D)\Lambda^{-1}=H(-\bm{k},-D)$. Second, this also means that, at $D=0$ the gyrotropy is zero. Third, a small displacement field of $\sim 0.01$ V/nm is sufficient to reach the maximum gyrotropy, so only weak symmetry breaking is necessary. Fourth, there is a crossover in the sign of the response at $\hbar\omega\sim 10$-$13$ meV: this originates from the intraband part of Eq. \ref{eq:kubo} dominating at small $\omega$ while the interband part dominates at large $\omega$ and they enter with different signs. Fifth, the decoupled extrinsic gyrotropy dominates the intrinsic electronic gyrotropy due to interlayer coupling since the $t_\perp=-8.5$ meV response is essentially the decoupled $t_\perp=0$ response with slight modifications.
            Together these results show that gyrotropy is both achievable in twisted bilayer MoTe$_2$ and useful as a diagnostic of crystalline symmetry breaking.

    \section{Outlook}

        In this paper we have emphasized that there are generically two contributions to optical gyrotropy in twisted bilayers: one intrinsic contribution due to interlayer electronic coupling and one due to the extrinsic geometry of the real-space embedding of the layers. While the former might be anticipated to always dominate, we have shown both in a toy model of a classical isolated bilayer wire array and a realistic model of twisted bilayer MoTe$_2$ with heterostrain and a displacement field that the extrinsic geometric contribution dominates. We analyzed symmetries with a focus on those relevant for moir\'e materials and identified conditions under which the gyrotropy is forbidden, purely intrinsic, or purely extrinsic.

        To focus on the essential physics we omitted electronic interactions, but considering the correlated electron regime and the effects on gyrotropy could be a fruitful direction. As the presence of gyrotropy is constrained by crystalline, pseudospin, and antiunitary symmetries it could serve as a sensitive probe of symmetry breaking. This could either be in twisted bilayers like we considered here, or more extended twisted stacks such as rhombohedral graphene where fractional Chern insulators were recently observed \cite{lu2024fractional}. The promise of these thicker twisted materials has already been realized for nonlinear optics \cite{kim2024three,ji2024opto,song2024observation,de2026high} and polarization control \cite{khaliji2022twisted}. Truly extended twisted crystals are a particularly interesting frontier for the gyrotropic effects as curvature and torsion are enriched in three dimensions, expanding the possibility for real-space embedding-dependent gyrotropy. The band structures \cite{tung2017origin,wu2020three}, electronic states \cite{wang2025decomposing,phong2025squeezing,park2026magnetoplasmons}, and electronic transport \cite{kazinski2022scattering,tani2023perpendicular} of three-dimensional twisted stacks have been considered, but their interface with optics and the resulting conductivities are less well understood. Alternatively considering the competition between magnetic circular dichroism and gyrotropy due to electronic coupling \cite{crosse2021faraday,mead2025terahertz}, the effects of strong electric fields \cite{kelardeh2014wannier,iafrate2020bloch,debeule2024floquet}, disorder \cite{joy2020transparent,talkington2025weak}, or non-equilibrium processes in bilayer graphene \cite{ollier2023energy,talkington2024linear,esparza2025exceptional}, and more exotic platforms where flat bands occur \cite{rhim2021singular,talkington2022dissipation,regnault2022catalogue} could clarify particularly tunable physics present in two-dimensional systems. Explicitly, gyrotropy is relevant in trilayers \cite{margetis2024optical,xiao2026interlayer} and Wang \textit{et al} constructed a general framework where gyrotropy can emerge in 3D systems from 2D layers \cite{wang2020optical}. Alternatively, measuring the gyrotropy in response to spatially inhomogeneous electric field distributions and light with orbital angular momentum can further probe the structure of electronic states \cite{yokoshi2026optical}.
        In conclusion, gyrotropy can be a sensitive probe of electronic and geometric physics in twisted systems: contributions from electronic physics and physical geometry must be carefully disentangled.

    \section*{Data Availability}

        All data and analysis code supporting the findings of this study are publicly available as \href{https://doi.org/10.5281/zenodo.21399024}{Zenodo: 21399024}.
    
    \section*{Acknowledgments}
        
        We thank C. De Beule and B. F. Mead for discussions on related topics, and thank W. T. Tai for comments on the manuscript. S. T. acknowledges support from the NSF under Grant No. DGE-1845298. This work was initiated by E. J. M. under grant DE-FG02-84ER45118 from the  U. S. Department of Energy.


\begin{thebibliography}{108}%
\makeatletter
\providecommand \@ifxundefined [1]{%
 \@ifx{#1\undefined}
}%
\providecommand \@ifnum [1]{%
 \ifnum #1\expandafter \@firstoftwo
 \else \expandafter \@secondoftwo
 \fi
}%
\providecommand \@ifx [1]{%
 \ifx #1\expandafter \@firstoftwo
 \else \expandafter \@secondoftwo
 \fi
}%
\providecommand \natexlab [1]{#1}%
\providecommand \enquote  [1]{``#1''}%
\providecommand \bibnamefont  [1]{#1}%
\providecommand \bibfnamefont [1]{#1}%
\providecommand \citenamefont [1]{#1}%
\providecommand \href@noop [0]{\@secondoftwo}%
\providecommand \href [0]{\begingroup \@sanitize@url \@href}%
\providecommand \@href[1]{\@@startlink{#1}\@@href}%
\providecommand \@@href[1]{\endgroup#1\@@endlink}%
\providecommand \@sanitize@url [0]{\catcode `\\12\catcode `\$12\catcode `\&12\catcode `\#12\catcode `\^12\catcode `\_12\catcode `\%12\relax}%
\providecommand \@@startlink[1]{}%
\providecommand \@@endlink[0]{}%
\providecommand \url  [0]{\begingroup\@sanitize@url \@url }%
\providecommand \@url [1]{\endgroup\@href {#1}{\urlprefix }}%
\providecommand \urlprefix  [0]{URL }%
\providecommand \Eprint [0]{\href }%
\providecommand \doibase [0]{https://doi.org/}%
\providecommand \selectlanguage [0]{\@gobble}%
\providecommand \bibinfo  [0]{\@secondoftwo}%
\providecommand \bibfield  [0]{\@secondoftwo}%
\providecommand \translation [1]{[#1]}%
\providecommand \BibitemOpen [0]{}%
\providecommand \bibitemStop [0]{}%
\providecommand \bibitemNoStop [0]{.\EOS\space}%
\providecommand \EOS [0]{\spacefactor3000\relax}%
\providecommand \BibitemShut  [1]{\csname bibitem#1\endcsname}%
\let\auto@bib@innerbib\@empty
\bibitem [{\citenamefont {Suarez-Morell}\ \emph {et~al.}(2010)\citenamefont {Suarez-Morell}, \citenamefont {Correa}, \citenamefont {Vargas}, \citenamefont {Pacheco},\ and\ \citenamefont {Barticevic}}]{morell2010flat}%
  \BibitemOpen
  \bibfield  {author} {\bibinfo {author} {\bibfnamefont {E.}~\bibnamefont {Suarez-Morell}}, \bibinfo {author} {\bibfnamefont {J.~D.}\ \bibnamefont {Correa}}, \bibinfo {author} {\bibfnamefont {P.}~\bibnamefont {Vargas}}, \bibinfo {author} {\bibfnamefont {M.}~\bibnamefont {Pacheco}},\ and\ \bibinfo {author} {\bibfnamefont {Z.}~\bibnamefont {Barticevic}},\ }\bibfield  {title} {\bibinfo {title} {{Flat bands in slightly twisted bilayer graphene: Tight-binding calculations}},\ }\href {https://doi.org/10.1103/PhysRevB.82.121407} {\bibfield  {journal} {\bibinfo  {journal} {Phys. Rev. B}\ }\textbf {\bibinfo {volume} {82}},\ \bibinfo {pages} {121407} (\bibinfo {year} {2010})}\BibitemShut {NoStop}%
\bibitem [{\citenamefont {de~Laissardiere}\ \emph {et~al.}(2010)\citenamefont {de~Laissardiere}, \citenamefont {Mayou},\ and\ \citenamefont {Magaud}}]{delaissardiere2010localization}%
  \BibitemOpen
  \bibfield  {author} {\bibinfo {author} {\bibfnamefont {G.~T.}\ \bibnamefont {de~Laissardiere}}, \bibinfo {author} {\bibfnamefont {D.}~\bibnamefont {Mayou}},\ and\ \bibinfo {author} {\bibfnamefont {L.}~\bibnamefont {Magaud}},\ }\bibfield  {title} {\bibinfo {title} {{Localization of Dirac electrons in rotated graphene bilayers}},\ }\href {https://doi.org/10.1021/nl902948m} {\bibfield  {journal} {\bibinfo  {journal} {Nano Lett.}\ }\textbf {\bibinfo {volume} {10}},\ \bibinfo {pages} {804} (\bibinfo {year} {2010})}\BibitemShut {NoStop}%
\bibitem [{\citenamefont {Bistritzer}\ and\ \citenamefont {MacDonald}(2011)}]{bistritzer2011moire}%
  \BibitemOpen
  \bibfield  {author} {\bibinfo {author} {\bibfnamefont {R.}~\bibnamefont {Bistritzer}}\ and\ \bibinfo {author} {\bibfnamefont {A.~H.}\ \bibnamefont {MacDonald}},\ }\bibfield  {title} {\bibinfo {title} {{Moire bands in twisted double-layer graphene}},\ }\href {https://doi.org/10.1073/pnas.1108174108} {\bibfield  {journal} {\bibinfo  {journal} {Proc. Natl. Acad. Sci. U.S.A.}\ }\textbf {\bibinfo {volume} {108}},\ \bibinfo {pages} {12233} (\bibinfo {year} {2011})}\BibitemShut {NoStop}%
\bibitem [{\citenamefont {Cao}\ \emph {et~al.}(2018{\natexlab{a}})\citenamefont {Cao}, \citenamefont {Fatemi}, \citenamefont {Fang}, \citenamefont {Watanabe}, \citenamefont {Taniguchi}, \citenamefont {Kaxiras},\ and\ \citenamefont {Jarillo-Herrero}}]{cao2018unconventional}%
  \BibitemOpen
  \bibfield  {author} {\bibinfo {author} {\bibfnamefont {Y.}~\bibnamefont {Cao}}, \bibinfo {author} {\bibfnamefont {V.}~\bibnamefont {Fatemi}}, \bibinfo {author} {\bibfnamefont {S.}~\bibnamefont {Fang}}, \bibinfo {author} {\bibfnamefont {K.}~\bibnamefont {Watanabe}}, \bibinfo {author} {\bibfnamefont {T.}~\bibnamefont {Taniguchi}}, \bibinfo {author} {\bibfnamefont {E.}~\bibnamefont {Kaxiras}},\ and\ \bibinfo {author} {\bibfnamefont {P.}~\bibnamefont {Jarillo-Herrero}},\ }\bibfield  {title} {\bibinfo {title} {{Unconventional superconductivity in magic-angle graphene superlattices}},\ }\href {https://doi.org/10.1038/nature26160} {\bibfield  {journal} {\bibinfo  {journal} {Nature}\ }\textbf {\bibinfo {volume} {556}},\ \bibinfo {pages} {43} (\bibinfo {year} {2018}{\natexlab{a}})}\BibitemShut {NoStop}%
\bibitem [{\citenamefont {Yankowitz}\ \emph {et~al.}(2019)\citenamefont {Yankowitz}, \citenamefont {Chen}, \citenamefont {Polshyn}, \citenamefont {Zhang}, \citenamefont {Watanabe}, \citenamefont {Taniguchi}, \citenamefont {Graf}, \citenamefont {Young},\ and\ \citenamefont {Dean}}]{yankowitz2019tuning}%
  \BibitemOpen
  \bibfield  {author} {\bibinfo {author} {\bibfnamefont {M.}~\bibnamefont {Yankowitz}}, \bibinfo {author} {\bibfnamefont {S.}~\bibnamefont {Chen}}, \bibinfo {author} {\bibfnamefont {H.}~\bibnamefont {Polshyn}}, \bibinfo {author} {\bibfnamefont {Y.}~\bibnamefont {Zhang}}, \bibinfo {author} {\bibfnamefont {K.}~\bibnamefont {Watanabe}}, \bibinfo {author} {\bibfnamefont {T.}~\bibnamefont {Taniguchi}}, \bibinfo {author} {\bibfnamefont {D.}~\bibnamefont {Graf}}, \bibinfo {author} {\bibfnamefont {A.~F.}\ \bibnamefont {Young}},\ and\ \bibinfo {author} {\bibfnamefont {C.~R.}\ \bibnamefont {Dean}},\ }\bibfield  {title} {\bibinfo {title} {{Tuning superconductivity in twisted bilayer graphene}},\ }\href {https://doi.org/10.1126/science.aav1910} {\bibfield  {journal} {\bibinfo  {journal} {Science}\ }\textbf {\bibinfo {volume} {363}},\ \bibinfo {pages} {1059} (\bibinfo {year} {2019})}\BibitemShut {NoStop}%
\bibitem [{\citenamefont {Cao}\ \emph {et~al.}(2018{\natexlab{b}})\citenamefont {Cao}, \citenamefont {Fatemi}, \citenamefont {Demir}, \citenamefont {Fang}, \citenamefont {Tomarken}, \citenamefont {Luo}, \citenamefont {Sanchez~Yamagishi}, \citenamefont {Watanabe}, \citenamefont {Taniguchi}, \citenamefont {Kaxiras} \emph {et~al.}}]{cao2018correlated}%
  \BibitemOpen
  \bibfield  {author} {\bibinfo {author} {\bibfnamefont {Y.}~\bibnamefont {Cao}}, \bibinfo {author} {\bibfnamefont {V.}~\bibnamefont {Fatemi}}, \bibinfo {author} {\bibfnamefont {A.}~\bibnamefont {Demir}}, \bibinfo {author} {\bibfnamefont {S.}~\bibnamefont {Fang}}, \bibinfo {author} {\bibfnamefont {S.~L.}\ \bibnamefont {Tomarken}}, \bibinfo {author} {\bibfnamefont {J.~Y.}\ \bibnamefont {Luo}}, \bibinfo {author} {\bibfnamefont {J.~D.}\ \bibnamefont {Sanchez~Yamagishi}}, \bibinfo {author} {\bibfnamefont {K.}~\bibnamefont {Watanabe}}, \bibinfo {author} {\bibfnamefont {T.}~\bibnamefont {Taniguchi}}, \bibinfo {author} {\bibfnamefont {E.}~\bibnamefont {Kaxiras}}, \emph {et~al.},\ }\bibfield  {title} {\bibinfo {title} {{Correlated insulator behaviour at half-filling in magic-angle graphene superlattices}},\ }\href {https://doi.org/10.1038/Nature26154} {\bibfield  {journal} {\bibinfo  {journal} {Nature}\ }\textbf {\bibinfo {volume} {556}},\ \bibinfo {pages} {80} (\bibinfo {year} {2018}{\natexlab{b}})}\BibitemShut
  {NoStop}%
\bibitem [{\citenamefont {Cao}\ \emph {et~al.}(2020)\citenamefont {Cao}, \citenamefont {Chowdhury}, \citenamefont {Rodan-Legrain}, \citenamefont {Rubies-Bigorda}, \citenamefont {Watanabe}, \citenamefont {Taniguchi}, \citenamefont {Senthil},\ and\ \citenamefont {Jarillo-Herrero}}]{cao2020strange}%
  \BibitemOpen
  \bibfield  {author} {\bibinfo {author} {\bibfnamefont {Y.}~\bibnamefont {Cao}}, \bibinfo {author} {\bibfnamefont {D.}~\bibnamefont {Chowdhury}}, \bibinfo {author} {\bibfnamefont {D.}~\bibnamefont {Rodan-Legrain}}, \bibinfo {author} {\bibfnamefont {O.}~\bibnamefont {Rubies-Bigorda}}, \bibinfo {author} {\bibfnamefont {K.}~\bibnamefont {Watanabe}}, \bibinfo {author} {\bibfnamefont {T.}~\bibnamefont {Taniguchi}}, \bibinfo {author} {\bibfnamefont {T.}~\bibnamefont {Senthil}},\ and\ \bibinfo {author} {\bibfnamefont {P.}~\bibnamefont {Jarillo-Herrero}},\ }\bibfield  {title} {\bibinfo {title} {{Strange metal in magic-angle graphene with near Planckian dissipation}},\ }\href {https://doi.org/10.1103/PhysRevLett.124.076801} {\bibfield  {journal} {\bibinfo  {journal} {Phys. Rev. Lett.}\ }\textbf {\bibinfo {volume} {124}},\ \bibinfo {pages} {076801} (\bibinfo {year} {2020})}\BibitemShut {NoStop}%
\bibitem [{\citenamefont {Wang}\ \emph {et~al.}(2020{\natexlab{a}})\citenamefont {Wang}, \citenamefont {Shih}, \citenamefont {Ghiotto}, \citenamefont {Xian}, \citenamefont {Rhodes}, \citenamefont {Tan}, \citenamefont {Claassen}, \citenamefont {Kennes}, \citenamefont {Bai}, \citenamefont {Kim} \emph {et~al.}}]{wang2019magic}%
  \BibitemOpen
  \bibfield  {author} {\bibinfo {author} {\bibfnamefont {L.}~\bibnamefont {Wang}}, \bibinfo {author} {\bibfnamefont {E.~M.}\ \bibnamefont {Shih}}, \bibinfo {author} {\bibfnamefont {A.}~\bibnamefont {Ghiotto}}, \bibinfo {author} {\bibfnamefont {L.}~\bibnamefont {Xian}}, \bibinfo {author} {\bibfnamefont {D.~A.}\ \bibnamefont {Rhodes}}, \bibinfo {author} {\bibfnamefont {C.}~\bibnamefont {Tan}}, \bibinfo {author} {\bibfnamefont {M.}~\bibnamefont {Claassen}}, \bibinfo {author} {\bibfnamefont {D.~M.}\ \bibnamefont {Kennes}}, \bibinfo {author} {\bibfnamefont {Y.}~\bibnamefont {Bai}}, \bibinfo {author} {\bibfnamefont {B.}~\bibnamefont {Kim}}, \emph {et~al.},\ }\bibfield  {title} {\bibinfo {title} {{Correlated electronic phases in twisted bilayer transition metal dichalcogenides}},\ }\href {https://doi.org/10.1038/s41563-020-0708-6} {\bibfield  {journal} {\bibinfo  {journal} {Nat. Mater.}\ }\textbf {\bibinfo {volume} {19}},\ \bibinfo {pages} {861} (\bibinfo {year} {2020}{\natexlab{a}})}\BibitemShut {NoStop}%
\bibitem [{\citenamefont {Devakul}\ \emph {et~al.}(2021)\citenamefont {Devakul}, \citenamefont {Crepel}, \citenamefont {Zhang},\ and\ \citenamefont {Fu}}]{devakul2021magic}%
  \BibitemOpen
  \bibfield  {author} {\bibinfo {author} {\bibfnamefont {T.}~\bibnamefont {Devakul}}, \bibinfo {author} {\bibfnamefont {V.}~\bibnamefont {Crepel}}, \bibinfo {author} {\bibfnamefont {Y.}~\bibnamefont {Zhang}},\ and\ \bibinfo {author} {\bibfnamefont {L.}~\bibnamefont {Fu}},\ }\bibfield  {title} {\bibinfo {title} {{Magic in twisted transition metal dichalcogenide bilayers}},\ }\href {https://doi.org/10.1038/s41467-021-27042-9} {\bibfield  {journal} {\bibinfo  {journal} {Nat. Commun.}\ }\textbf {\bibinfo {volume} {12}},\ \bibinfo {pages} {6730} (\bibinfo {year} {2021})}\BibitemShut {NoStop}%
\bibitem [{\citenamefont {Andrei}\ \emph {et~al.}(2021)\citenamefont {Andrei}, \citenamefont {Efetov}, \citenamefont {Jarillo-Herrero}, \citenamefont {MacDonald}, \citenamefont {Mak}, \citenamefont {Senthil}, \citenamefont {Tutuc}, \citenamefont {Yazdani},\ and\ \citenamefont {Young}}]{andrei2021marvels}%
  \BibitemOpen
  \bibfield  {author} {\bibinfo {author} {\bibfnamefont {E.~Y.}\ \bibnamefont {Andrei}}, \bibinfo {author} {\bibfnamefont {D.~K.}\ \bibnamefont {Efetov}}, \bibinfo {author} {\bibfnamefont {P.}~\bibnamefont {Jarillo-Herrero}}, \bibinfo {author} {\bibfnamefont {A.~H.}\ \bibnamefont {MacDonald}}, \bibinfo {author} {\bibfnamefont {K.~F.}\ \bibnamefont {Mak}}, \bibinfo {author} {\bibfnamefont {T.}~\bibnamefont {Senthil}}, \bibinfo {author} {\bibfnamefont {E.}~\bibnamefont {Tutuc}}, \bibinfo {author} {\bibfnamefont {A.}~\bibnamefont {Yazdani}},\ and\ \bibinfo {author} {\bibfnamefont {A.~F.}\ \bibnamefont {Young}},\ }\bibfield  {title} {\bibinfo {title} {{The marvels of moire materials}},\ }\href {https://doi.org/10.1038/s41578-021-00284-1} {\bibfield  {journal} {\bibinfo  {journal} {Nat. Rev. Mater.}\ }\textbf {\bibinfo {volume} {6}},\ \bibinfo {pages} {201} (\bibinfo {year} {2021})}\BibitemShut {NoStop}%
\bibitem [{\citenamefont {Suarez-Morell}\ and\ \citenamefont {Torres}(2012)}]{morell2012radiation}%
  \BibitemOpen
  \bibfield  {author} {\bibinfo {author} {\bibfnamefont {E.}~\bibnamefont {Suarez-Morell}}\ and\ \bibinfo {author} {\bibfnamefont {L.~E.~F.}\ \bibnamefont {Torres}},\ }\bibfield  {title} {\bibinfo {title} {{Radiation effects on the electronic properties of bilayer graphene}},\ }\href {https://doi.org/10.1103/PhysRevB.86.125449} {\bibfield  {journal} {\bibinfo  {journal} {Phys. Rev. B}\ }\textbf {\bibinfo {volume} {86}},\ \bibinfo {pages} {125449} (\bibinfo {year} {2012})}\BibitemShut {NoStop}%
\bibitem [{\citenamefont {Tabert}\ and\ \citenamefont {Nicol}(2013)}]{tabert2013optical}%
  \BibitemOpen
  \bibfield  {author} {\bibinfo {author} {\bibfnamefont {C.~J.}\ \bibnamefont {Tabert}}\ and\ \bibinfo {author} {\bibfnamefont {E.~J.}\ \bibnamefont {Nicol}},\ }\bibfield  {title} {\bibinfo {title} {{Optical conductivity of twisted bilayer graphene}},\ }\href {https://doi.org/10.1103/PhysRevB.87.121402} {\bibfield  {journal} {\bibinfo  {journal} {Phys. Rev. B}\ }\textbf {\bibinfo {volume} {87}},\ \bibinfo {pages} {121402} (\bibinfo {year} {2013})}\BibitemShut {NoStop}%
\bibitem [{\citenamefont {Stauber}\ \emph {et~al.}(2013)\citenamefont {Stauber}, \citenamefont {San-Jose},\ and\ \citenamefont {Brey}}]{stauber2013optical}%
  \BibitemOpen
  \bibfield  {author} {\bibinfo {author} {\bibfnamefont {T.}~\bibnamefont {Stauber}}, \bibinfo {author} {\bibfnamefont {P.}~\bibnamefont {San-Jose}},\ and\ \bibinfo {author} {\bibfnamefont {L.}~\bibnamefont {Brey}},\ }\bibfield  {title} {\bibinfo {title} {{Optical conductivity, Drude weight and plasmons in twisted graphene bilayers}},\ }\href {https://doi.org/10.1088/1367-2630/15/11/113050} {\bibfield  {journal} {\bibinfo  {journal} {New J. Phys.}\ }\textbf {\bibinfo {volume} {15}},\ \bibinfo {pages} {113050} (\bibinfo {year} {2013})}\BibitemShut {NoStop}%
\bibitem [{\citenamefont {Moon}\ and\ \citenamefont {Koshino}(2013)}]{moon2013optical}%
  \BibitemOpen
  \bibfield  {author} {\bibinfo {author} {\bibfnamefont {P.}~\bibnamefont {Moon}}\ and\ \bibinfo {author} {\bibfnamefont {M.}~\bibnamefont {Koshino}},\ }\bibfield  {title} {\bibinfo {title} {{Optical absorption in twisted bilayer graphene}},\ }\href {https://doi.org/10.1103/PhysRevB.87.205404} {\bibfield  {journal} {\bibinfo  {journal} {Phys. Rev. B}\ }\textbf {\bibinfo {volume} {87}},\ \bibinfo {pages} {205404} (\bibinfo {year} {2013})}\BibitemShut {NoStop}%
\bibitem [{\citenamefont {Stauber}\ \emph {et~al.}(2023)\citenamefont {Stauber}, \citenamefont {Wackerl}, \citenamefont {Wenk}, \citenamefont {Margetis}, \citenamefont {Gonzalez}, \citenamefont {Gomez-Santos},\ and\ \citenamefont {Schliemann}}]{stauber2023neutral}%
  \BibitemOpen
  \bibfield  {author} {\bibinfo {author} {\bibfnamefont {T.}~\bibnamefont {Stauber}}, \bibinfo {author} {\bibfnamefont {M.}~\bibnamefont {Wackerl}}, \bibinfo {author} {\bibfnamefont {P.}~\bibnamefont {Wenk}}, \bibinfo {author} {\bibfnamefont {D.}~\bibnamefont {Margetis}}, \bibinfo {author} {\bibfnamefont {J.}~\bibnamefont {Gonzalez}}, \bibinfo {author} {\bibfnamefont {G.}~\bibnamefont {Gomez-Santos}},\ and\ \bibinfo {author} {\bibfnamefont {J.}~\bibnamefont {Schliemann}},\ }\bibfield  {title} {\bibinfo {title} {Neutral magic-angle bilayer graphene: Condon instability and chiral resonances},\ }\href {https://doi.org/10.1002/smsc.202200080} {\bibfield  {journal} {\bibinfo  {journal} {Small Science}\ }\textbf {\bibinfo {volume} {3}},\ \bibinfo {pages} {2200080} (\bibinfo {year} {2023})}\BibitemShut {NoStop}%
\bibitem [{\citenamefont {Kim}\ \emph {et~al.}(2016)\citenamefont {Kim}, \citenamefont {Sanchez-Castillo}, \citenamefont {Ziegler}, \citenamefont {Ogawa}, \citenamefont {Noguez},\ and\ \citenamefont {Park}}]{kim2016chiral}%
  \BibitemOpen
  \bibfield  {author} {\bibinfo {author} {\bibfnamefont {C.~J.}\ \bibnamefont {Kim}}, \bibinfo {author} {\bibfnamefont {A.}~\bibnamefont {Sanchez-Castillo}}, \bibinfo {author} {\bibfnamefont {Z.}~\bibnamefont {Ziegler}}, \bibinfo {author} {\bibfnamefont {Y.}~\bibnamefont {Ogawa}}, \bibinfo {author} {\bibfnamefont {C.}~\bibnamefont {Noguez}},\ and\ \bibinfo {author} {\bibfnamefont {J.}~\bibnamefont {Park}},\ }\bibfield  {title} {\bibinfo {title} {{Chiral atomically thin films}},\ }\href {https://doi.org/10.1038/NNANO.2016.3} {\bibfield  {journal} {\bibinfo  {journal} {Nat. Nanotechnol.}\ }\textbf {\bibinfo {volume} {11}},\ \bibinfo {pages} {520} (\bibinfo {year} {2016})}\BibitemShut {NoStop}%
\bibitem [{\citenamefont {Suarez-Morell}\ \emph {et~al.}(2017)\citenamefont {Suarez-Morell}, \citenamefont {Chico},\ and\ \citenamefont {Brey}}]{morell2017twisting}%
  \BibitemOpen
  \bibfield  {author} {\bibinfo {author} {\bibfnamefont {E.}~\bibnamefont {Suarez-Morell}}, \bibinfo {author} {\bibfnamefont {L.}~\bibnamefont {Chico}},\ and\ \bibinfo {author} {\bibfnamefont {L.}~\bibnamefont {Brey}},\ }\bibfield  {title} {\bibinfo {title} {{Twisting {D}irac fermions: circular dichroism in bilayer graphene}},\ }\href {https://doi.org/10.1088/2053-1583/aa7eb6} {\bibfield  {journal} {\bibinfo  {journal} {2D Mater.}\ }\textbf {\bibinfo {volume} {4}},\ \bibinfo {pages} {035015} (\bibinfo {year} {2017})}\BibitemShut {NoStop}%
\bibitem [{\citenamefont {Stauber}\ \emph {et~al.}(2018{\natexlab{a}})\citenamefont {Stauber}, \citenamefont {Low},\ and\ \citenamefont {Gomez-Santos}}]{stauber2018chiral}%
  \BibitemOpen
  \bibfield  {author} {\bibinfo {author} {\bibfnamefont {T.}~\bibnamefont {Stauber}}, \bibinfo {author} {\bibfnamefont {T.}~\bibnamefont {Low}},\ and\ \bibinfo {author} {\bibfnamefont {G.}~\bibnamefont {Gomez-Santos}},\ }\bibfield  {title} {\bibinfo {title} {{Chiral response of twisted bilayer graphene}},\ }\href {https://doi.org/10.1103/PhysRevLett.120.046801} {\bibfield  {journal} {\bibinfo  {journal} {Phys. Rev. Lett.}\ }\textbf {\bibinfo {volume} {120}},\ \bibinfo {pages} {046801} (\bibinfo {year} {2018}{\natexlab{a}})}\BibitemShut {NoStop}%
\bibitem [{\citenamefont {Stauber}\ \emph {et~al.}(2018{\natexlab{b}})\citenamefont {Stauber}, \citenamefont {Low},\ and\ \citenamefont {Gomez-Santos}}]{stauber2018linear}%
  \BibitemOpen
  \bibfield  {author} {\bibinfo {author} {\bibfnamefont {T.}~\bibnamefont {Stauber}}, \bibinfo {author} {\bibfnamefont {T.}~\bibnamefont {Low}},\ and\ \bibinfo {author} {\bibfnamefont {G.}~\bibnamefont {Gomez-Santos}},\ }\bibfield  {title} {\bibinfo {title} {{Linear response of twisted bilayer graphene: Continuum versus tight-binding models}},\ }\href {https://doi.org/10.1103/PhysRevB.98.195414} {\bibfield  {journal} {\bibinfo  {journal} {Phys. Rev. B}\ }\textbf {\bibinfo {volume} {98}},\ \bibinfo {pages} {195414} (\bibinfo {year} {2018}{\natexlab{b}})}\BibitemShut {NoStop}%
\bibitem [{\citenamefont {Addison}\ \emph {et~al.}(2019)\citenamefont {Addison}, \citenamefont {Park},\ and\ \citenamefont {Mele}}]{addison2019twist}%
  \BibitemOpen
  \bibfield  {author} {\bibinfo {author} {\bibfnamefont {Z.}~\bibnamefont {Addison}}, \bibinfo {author} {\bibfnamefont {J.}~\bibnamefont {Park}},\ and\ \bibinfo {author} {\bibfnamefont {E.~J.}\ \bibnamefont {Mele}},\ }\bibfield  {title} {\bibinfo {title} {{Twist, slip, and circular dichroism in bilayer graphene}},\ }\href {https://doi.org/10.1103/PhysRevB.100.125418} {\bibfield  {journal} {\bibinfo  {journal} {Phys. Rev. B}\ }\textbf {\bibinfo {volume} {100}},\ \bibinfo {pages} {125418} (\bibinfo {year} {2019})}\BibitemShut {NoStop}%
\bibitem [{\citenamefont {Do}\ \emph {et~al.}(2020)\citenamefont {Do}, \citenamefont {Le}, \citenamefont {Nguyen},\ and\ \citenamefont {Bercioux}}]{do2020optical}%
  \BibitemOpen
  \bibfield  {author} {\bibinfo {author} {\bibfnamefont {V.~N.}\ \bibnamefont {Do}}, \bibinfo {author} {\bibfnamefont {H.~A.}\ \bibnamefont {Le}}, \bibinfo {author} {\bibfnamefont {V.~D.}\ \bibnamefont {Nguyen}},\ and\ \bibinfo {author} {\bibfnamefont {D.}~\bibnamefont {Bercioux}},\ }\bibfield  {title} {\bibinfo {title} {{Optical Hall response of bilayer graphene: Manifestation of chiral hybridized states in broken mirror symmetry lattices}},\ }\href {https://doi.org/10.1103/PhysRevResearch.2.043281} {\bibfield  {journal} {\bibinfo  {journal} {Phys. Rev. Res.}\ }\textbf {\bibinfo {volume} {2}},\ \bibinfo {pages} {043281} (\bibinfo {year} {2020})}\BibitemShut {NoStop}%
\bibitem [{\citenamefont {Ho}\ and\ \citenamefont {Do}(2023)}]{ho2023optical}%
  \BibitemOpen
  \bibfield  {author} {\bibinfo {author} {\bibfnamefont {S.~T.}\ \bibnamefont {Ho}}\ and\ \bibinfo {author} {\bibfnamefont {V.~N.}\ \bibnamefont {Do}},\ }\bibfield  {title} {\bibinfo {title} {{Optical activity and transport in twisted bilayer graphene: Spatial dispersion effects}},\ }\href {https://doi.org/10.1103/PhysRevB.107.195141} {\bibfield  {journal} {\bibinfo  {journal} {Phys. Rev. B}\ }\textbf {\bibinfo {volume} {107}},\ \bibinfo {pages} {195141} (\bibinfo {year} {2023})}\BibitemShut {NoStop}%
\bibitem [{\citenamefont {Qiu}\ \emph {et~al.}(2024)\citenamefont {Qiu}, \citenamefont {Yang}, \citenamefont {Chu},\ and\ \citenamefont {Yan}}]{qiu2024detection}%
  \BibitemOpen
  \bibfield  {author} {\bibinfo {author} {\bibfnamefont {X.~M.}\ \bibnamefont {Qiu}}, \bibinfo {author} {\bibfnamefont {N.}~\bibnamefont {Yang}}, \bibinfo {author} {\bibfnamefont {W.}~\bibnamefont {Chu}},\ and\ \bibinfo {author} {\bibfnamefont {J.~Y.}\ \bibnamefont {Yan}},\ }\bibfield  {title} {\bibinfo {title} {{Detection of the chirality of twisted bilayer graphene by the optical absorption}},\ }\href {https://doi.org/10.1103/PhysRevB.109.125419} {\bibfield  {journal} {\bibinfo  {journal} {Phys. Rev. B}\ }\textbf {\bibinfo {volume} {109}},\ \bibinfo {pages} {125419} (\bibinfo {year} {2024})}\BibitemShut {NoStop}%
\bibitem [{\citenamefont {Shallcross}\ \emph {et~al.}(2008)\citenamefont {Shallcross}, \citenamefont {Sharma},\ and\ \citenamefont {Pankratov}}]{shallcross2008quantum}%
  \BibitemOpen
  \bibfield  {author} {\bibinfo {author} {\bibfnamefont {S.}~\bibnamefont {Shallcross}}, \bibinfo {author} {\bibfnamefont {S.}~\bibnamefont {Sharma}},\ and\ \bibinfo {author} {\bibfnamefont {O.~A.}\ \bibnamefont {Pankratov}},\ }\bibfield  {title} {\bibinfo {title} {{Quantum interference at the twist boundary in graphene}},\ }\href {https://doi.org/10.1103/PhysRevLett.101.056803} {\bibfield  {journal} {\bibinfo  {journal} {Phys. Rev. Lett.}\ }\textbf {\bibinfo {volume} {101}},\ \bibinfo {pages} {056803} (\bibinfo {year} {2008})}\BibitemShut {NoStop}%
\bibitem [{\citenamefont {Mele}(2010)}]{mele2010commensuration}%
  \BibitemOpen
  \bibfield  {author} {\bibinfo {author} {\bibfnamefont {E.~J.}\ \bibnamefont {Mele}},\ }\bibfield  {title} {\bibinfo {title} {{Commensuration and interlayer coherence in twisted bilayer graphene}},\ }\href {https://doi.org/10.1103/PhysRevB.81.161405} {\bibfield  {journal} {\bibinfo  {journal} {Phys. Rev. B}\ }\textbf {\bibinfo {volume} {81}},\ \bibinfo {pages} {161405} (\bibinfo {year} {2010})}\BibitemShut {NoStop}%
\bibitem [{\citenamefont {Mele}(2012)}]{mele2012interlayer}%
  \BibitemOpen
  \bibfield  {author} {\bibinfo {author} {\bibfnamefont {E.~J.}\ \bibnamefont {Mele}},\ }\bibfield  {title} {\bibinfo {title} {{Interlayer coupling in rotationally faulted multilayer graphenes}},\ }\href {https://doi.org/10.1088/0022-3727/45/15/154004} {\bibfield  {journal} {\bibinfo  {journal} {J. Phys. D: Appl. Phys.}\ }\textbf {\bibinfo {volume} {45}},\ \bibinfo {pages} {154004} (\bibinfo {year} {2012})}\BibitemShut {NoStop}%
\bibitem [{\citenamefont {Talkington}\ and\ \citenamefont {Mele}(2023{\natexlab{a}})}]{talkington2023electric}%
  \BibitemOpen
  \bibfield  {author} {\bibinfo {author} {\bibfnamefont {S.}~\bibnamefont {Talkington}}\ and\ \bibinfo {author} {\bibfnamefont {E.~J.}\ \bibnamefont {Mele}},\ }\bibfield  {title} {\bibinfo {title} {{Electric-field-tunable band gap in commensurate twisted bilayer graphene}},\ }\href {https://doi.org/10.1103/PhysRevB.107.L041408} {\bibfield  {journal} {\bibinfo  {journal} {Phys. Rev. B}\ }\textbf {\bibinfo {volume} {107}},\ \bibinfo {pages} {L041408} (\bibinfo {year} {2023}{\natexlab{a}})}\BibitemShut {NoStop}%
\bibitem [{\citenamefont {Talkington}\ and\ \citenamefont {Mele}(2023{\natexlab{b}})}]{talkington2023terahertz}%
  \BibitemOpen
  \bibfield  {author} {\bibinfo {author} {\bibfnamefont {S.}~\bibnamefont {Talkington}}\ and\ \bibinfo {author} {\bibfnamefont {E.~J.}\ \bibnamefont {Mele}},\ }\bibfield  {title} {\bibinfo {title} {{Terahertz Circular Dichroism in Commensurate Twisted Bilayer Graphene}},\ }\href {https://doi.org/10.1103/PhysRevB.108.085421} {\bibfield  {journal} {\bibinfo  {journal} {Phys. Rev. B}\ }\textbf {\bibinfo {volume} {108}},\ \bibinfo {pages} {085421} (\bibinfo {year} {2023}{\natexlab{b}})}\BibitemShut {NoStop}%
\bibitem [{\citenamefont {Huang}\ \emph {et~al.}(2022)\citenamefont {Huang}, \citenamefont {Tu}, \citenamefont {Shen}, \citenamefont {Zheng}, \citenamefont {Wang}, \citenamefont {Wang}, \citenamefont {Khaliji}, \citenamefont {Park}, \citenamefont {Liu}, \citenamefont {Yang} \emph {et~al.}}]{huang2022observation}%
  \BibitemOpen
  \bibfield  {author} {\bibinfo {author} {\bibfnamefont {T.}~\bibnamefont {Huang}}, \bibinfo {author} {\bibfnamefont {X.}~\bibnamefont {Tu}}, \bibinfo {author} {\bibfnamefont {C.}~\bibnamefont {Shen}}, \bibinfo {author} {\bibfnamefont {B.}~\bibnamefont {Zheng}}, \bibinfo {author} {\bibfnamefont {J.}~\bibnamefont {Wang}}, \bibinfo {author} {\bibfnamefont {H.}~\bibnamefont {Wang}}, \bibinfo {author} {\bibfnamefont {K.}~\bibnamefont {Khaliji}}, \bibinfo {author} {\bibfnamefont {S.~H.}\ \bibnamefont {Park}}, \bibinfo {author} {\bibfnamefont {Z.}~\bibnamefont {Liu}}, \bibinfo {author} {\bibfnamefont {T.}~\bibnamefont {Yang}}, \emph {et~al.},\ }\bibfield  {title} {\bibinfo {title} {{Observation of chiral and slow plasmons in twisted bilayer graphene}},\ }\href {https://doi.org/10.1038/s41586-022-04520-8} {\bibfield  {journal} {\bibinfo  {journal} {Nature}\ }\textbf {\bibinfo {volume} {605}},\ \bibinfo {pages} {63} (\bibinfo {year} {2022})}\BibitemShut {NoStop}%
\bibitem [{\citenamefont {Lan}\ \emph {et~al.}(2021)\citenamefont {Lan}, \citenamefont {Liu}, \citenamefont {Wang}, \citenamefont {Zhu}, \citenamefont {Liu}, \citenamefont {Gong}, \citenamefont {Yang}, \citenamefont {Shi}, \citenamefont {Wang},\ and\ \citenamefont {Zhang}}]{lan2021observation}%
  \BibitemOpen
  \bibfield  {author} {\bibinfo {author} {\bibfnamefont {S.}~\bibnamefont {Lan}}, \bibinfo {author} {\bibfnamefont {X.}~\bibnamefont {Liu}}, \bibinfo {author} {\bibfnamefont {S.}~\bibnamefont {Wang}}, \bibinfo {author} {\bibfnamefont {H.}~\bibnamefont {Zhu}}, \bibinfo {author} {\bibfnamefont {Y.}~\bibnamefont {Liu}}, \bibinfo {author} {\bibfnamefont {C.}~\bibnamefont {Gong}}, \bibinfo {author} {\bibfnamefont {S.}~\bibnamefont {Yang}}, \bibinfo {author} {\bibfnamefont {J.}~\bibnamefont {Shi}}, \bibinfo {author} {\bibfnamefont {Y.}~\bibnamefont {Wang}},\ and\ \bibinfo {author} {\bibfnamefont {X.}~\bibnamefont {Zhang}},\ }\bibfield  {title} {\bibinfo {title} {{Observation of strong excitonic magneto-chiral anisotropy in twisted bilayer van der Waals crystals}},\ }\href {https://doi.org/10.1038/s41467-021-22412-9} {\bibfield  {journal} {\bibinfo  {journal} {Nat. Commun.}\ }\textbf {\bibinfo {volume} {12}},\ \bibinfo {pages} {2088} (\bibinfo {year} {2021})}\BibitemShut {NoStop}%
\bibitem [{\citenamefont {Kim}\ \emph {et~al.}(2024)\citenamefont {Kim}, \citenamefont {Jin}, \citenamefont {Wang}, \citenamefont {He}, \citenamefont {Christensen}, \citenamefont {Mele},\ and\ \citenamefont {Zhen}}]{kim2024three}%
  \BibitemOpen
  \bibfield  {author} {\bibinfo {author} {\bibfnamefont {B.}~\bibnamefont {Kim}}, \bibinfo {author} {\bibfnamefont {J.}~\bibnamefont {Jin}}, \bibinfo {author} {\bibfnamefont {Z.}~\bibnamefont {Wang}}, \bibinfo {author} {\bibfnamefont {L.}~\bibnamefont {He}}, \bibinfo {author} {\bibfnamefont {T.}~\bibnamefont {Christensen}}, \bibinfo {author} {\bibfnamefont {E.~J.}\ \bibnamefont {Mele}},\ and\ \bibinfo {author} {\bibfnamefont {B.}~\bibnamefont {Zhen}},\ }\bibfield  {title} {\bibinfo {title} {{Three-dimensional nonlinear optical materials from twisted two-dimensional van der Waals interfaces}},\ }\href {https://doi.org/10.1038/s41566-023-01318-6} {\bibfield  {journal} {\bibinfo  {journal} {Nat. Photonics}\ }\textbf {\bibinfo {volume} {18}},\ \bibinfo {pages} {91} (\bibinfo {year} {2024})}\BibitemShut {NoStop}%
\bibitem [{\citenamefont {Nguyen}\ and\ \citenamefont {Son}(2020)}]{nguyen2020electrodynamics}%
  \BibitemOpen
  \bibfield  {author} {\bibinfo {author} {\bibfnamefont {D.~X.}\ \bibnamefont {Nguyen}}\ and\ \bibinfo {author} {\bibfnamefont {D.~T.}\ \bibnamefont {Son}},\ }\bibfield  {title} {\bibinfo {title} {{Electrodynamics of thin sheets of twisted material}},\ }\bibfield  {journal} {\bibinfo  {journal} {arXiv}\ }\href {https://doi.org/10.48550/arXiv.2008.02812} {10.48550/arXiv.2008.02812} (\bibinfo {year} {2020})\BibitemShut {NoStop}%
\bibitem [{\citenamefont {Ochoa}\ and\ \citenamefont {Asenjo-Garcia}(2020)}]{ochoa2020flat}%
  \BibitemOpen
  \bibfield  {author} {\bibinfo {author} {\bibfnamefont {H.}~\bibnamefont {Ochoa}}\ and\ \bibinfo {author} {\bibfnamefont {A.}~\bibnamefont {Asenjo-Garcia}},\ }\bibfield  {title} {\bibinfo {title} {{Flat bands and chiral optical response of moire insulators}},\ }\href {https://doi.org/10.1103/PhysRevLett.125.037402} {\bibfield  {journal} {\bibinfo  {journal} {Phys. Rev. Lett.}\ }\textbf {\bibinfo {volume} {125}},\ \bibinfo {pages} {037402} (\bibinfo {year} {2020})}\BibitemShut {NoStop}%
\bibitem [{\citenamefont {Ding}\ and\ \citenamefont {Zhao}(2023)}]{ding2023chiral}%
  \BibitemOpen
  \bibfield  {author} {\bibinfo {author} {\bibfnamefont {C.}~\bibnamefont {Ding}}\ and\ \bibinfo {author} {\bibfnamefont {M.}~\bibnamefont {Zhao}},\ }\bibfield  {title} {\bibinfo {title} {{Chiral response in two-dimensional bilayers with time-reversal symmetry: A universal criterion}},\ }\href {https://doi.org/10.1103/PhysRevB.108.125415} {\bibfield  {journal} {\bibinfo  {journal} {Phys. Rev. B}\ }\textbf {\bibinfo {volume} {108}},\ \bibinfo {pages} {125415} (\bibinfo {year} {2023})}\BibitemShut {NoStop}%
\bibitem [{\citenamefont {Bistritzer}\ and\ \citenamefont {MacDonald}(2010)}]{bistritzer2010transport}%
  \BibitemOpen
  \bibfield  {author} {\bibinfo {author} {\bibfnamefont {R.}~\bibnamefont {Bistritzer}}\ and\ \bibinfo {author} {\bibfnamefont {A.~H.}\ \bibnamefont {MacDonald}},\ }\bibfield  {title} {\bibinfo {title} {{Transport between twisted graphene layers}},\ }\href {https://doi.org/10.1103/PhysRevB.81.245412} {\bibfield  {journal} {\bibinfo  {journal} {Phys. Rev. B}\ }\textbf {\bibinfo {volume} {81}},\ \bibinfo {pages} {245412} (\bibinfo {year} {2010})}\BibitemShut {NoStop}%
\bibitem [{\citenamefont {Zhu}\ \emph {et~al.}(2024)\citenamefont {Zhu}, \citenamefont {Zhai}, \citenamefont {Xiao},\ and\ \citenamefont {Yao}}]{zhu2024layer}%
  \BibitemOpen
  \bibfield  {author} {\bibinfo {author} {\bibfnamefont {J.}~\bibnamefont {Zhu}}, \bibinfo {author} {\bibfnamefont {D.}~\bibnamefont {Zhai}}, \bibinfo {author} {\bibfnamefont {C.}~\bibnamefont {Xiao}},\ and\ \bibinfo {author} {\bibfnamefont {W.}~\bibnamefont {Yao}},\ }\bibfield  {title} {\bibinfo {title} {{Layer Hall counterflow as a model probe of magic-angle twisted bilayer graphene}},\ }\href {https://doi.org/10.1103/PhysRevB.109.155114} {\bibfield  {journal} {\bibinfo  {journal} {Phys. Rev. B}\ }\textbf {\bibinfo {volume} {109}},\ \bibinfo {pages} {155114} (\bibinfo {year} {2024})}\BibitemShut {NoStop}%
\bibitem [{\citenamefont {Franta}(2020)}]{franta2020symmetry}%
  \BibitemOpen
  \bibfield  {author} {\bibinfo {author} {\bibfnamefont {D.}~\bibnamefont {Franta}},\ }\bibfield  {title} {\bibinfo {title} {{Symmetry of linear dielectric response tensors: dispersion models fulfilling three fundamental conditions}},\ }\href {https://doi.org/10.1063/5.0005735} {\bibfield  {journal} {\bibinfo  {journal} {J. Appl. Phys.}\ }\textbf {\bibinfo {volume} {127}},\ \bibinfo {pages} {223101} (\bibinfo {year} {2020})}\BibitemShut {NoStop}%
\bibitem [{\citenamefont {Pozo~Ocana}\ and\ \citenamefont {Souza}(2023)}]{pozo2023multipole}%
  \BibitemOpen
  \bibfield  {author} {\bibinfo {author} {\bibfnamefont {O.}~\bibnamefont {Pozo~Ocana}}\ and\ \bibinfo {author} {\bibfnamefont {I.}~\bibnamefont {Souza}},\ }\bibfield  {title} {\bibinfo {title} {{Multipole theory of optical spatial dispersion in crystals}},\ }\href {https://doi.org/10.21468/SciPostPhys.14.5.118} {\bibfield  {journal} {\bibinfo  {journal} {SciPost Phys.}\ }\textbf {\bibinfo {volume} {14}},\ \bibinfo {pages} {118} (\bibinfo {year} {2023})}\BibitemShut {NoStop}%
\bibitem [{\citenamefont {Avdoshkin}\ and\ \citenamefont {Popov}(2023)}]{avdoshkin2023extrinsic}%
  \BibitemOpen
  \bibfield  {author} {\bibinfo {author} {\bibfnamefont {A.}~\bibnamefont {Avdoshkin}}\ and\ \bibinfo {author} {\bibfnamefont {F.~K.}\ \bibnamefont {Popov}},\ }\bibfield  {title} {\bibinfo {title} {{Extrinsic geometry of quantum states}},\ }\href {https://doi.org/10.1103/PhysRevB.107.245136} {\bibfield  {journal} {\bibinfo  {journal} {Phys. Rev. B}\ }\textbf {\bibinfo {volume} {107}},\ \bibinfo {pages} {245136} (\bibinfo {year} {2023})}\BibitemShut {NoStop}%
\bibitem [{\citenamefont {Wang}\ \emph {et~al.}(2020{\natexlab{b}})\citenamefont {Wang}, \citenamefont {Morimoto},\ and\ \citenamefont {Moore}}]{wang2020optical}%
  \BibitemOpen
  \bibfield  {author} {\bibinfo {author} {\bibfnamefont {Y.~Q.}\ \bibnamefont {Wang}}, \bibinfo {author} {\bibfnamefont {T.}~\bibnamefont {Morimoto}},\ and\ \bibinfo {author} {\bibfnamefont {J.~E.}\ \bibnamefont {Moore}},\ }\bibfield  {title} {\bibinfo {title} {{Optical rotation in thin chiral/twisted materials and the gyrotropic magnetic effect}},\ }\href {https://doi.org/10.1103/PhysRevB.101.174419} {\bibfield  {journal} {\bibinfo  {journal} {Phys. Rev. B}\ }\textbf {\bibinfo {volume} {101}},\ \bibinfo {pages} {174419} (\bibinfo {year} {2020}{\natexlab{b}})}\BibitemShut {NoStop}%
\bibitem [{\citenamefont {Mahon}\ and\ \citenamefont {Sipe}(2020)}]{mahon2020magnetoelectric}%
  \BibitemOpen
  \bibfield  {author} {\bibinfo {author} {\bibfnamefont {P.~T.}\ \bibnamefont {Mahon}}\ and\ \bibinfo {author} {\bibfnamefont {J.~E.}\ \bibnamefont {Sipe}},\ }\bibfield  {title} {\bibinfo {title} {{From magnetoelectric response to optical activity}},\ }\href {https://doi.org/10.1103/PhysRevResearch.2.043110} {\bibfield  {journal} {\bibinfo  {journal} {Phys. Rev. Res.}\ }\textbf {\bibinfo {volume} {2}},\ \bibinfo {pages} {043110} (\bibinfo {year} {2020})}\BibitemShut {NoStop}%
\bibitem [{\citenamefont {Zou}\ \emph {et~al.}(2018)\citenamefont {Zou}, \citenamefont {Po}, \citenamefont {Vishwanath},\ and\ \citenamefont {Senthil}}]{zou2018band}%
  \BibitemOpen
  \bibfield  {author} {\bibinfo {author} {\bibfnamefont {L.}~\bibnamefont {Zou}}, \bibinfo {author} {\bibfnamefont {H.~C.}\ \bibnamefont {Po}}, \bibinfo {author} {\bibfnamefont {A.}~\bibnamefont {Vishwanath}},\ and\ \bibinfo {author} {\bibfnamefont {T.}~\bibnamefont {Senthil}},\ }\bibfield  {title} {\bibinfo {title} {{Band structure of twisted bilayer graphene: Emergent symmetries, commensurate approximants, and Wannier obstructions}},\ }\href {https://doi.org/10.1103/PhysRevB.98.085435} {\bibfield  {journal} {\bibinfo  {journal} {Phys. Rev. B}\ }\textbf {\bibinfo {volume} {98}},\ \bibinfo {pages} {085435} (\bibinfo {year} {2018})}\BibitemShut {NoStop}%
\bibitem [{\citenamefont {Yu}(2024)}]{yu2024strongly}%
  \BibitemOpen
  \bibfield  {author} {\bibinfo {author} {\bibfnamefont {G.}~\bibnamefont {Yu}},\ }\emph {\bibinfo {title} {Strongly Correlated Quantum States in WTe2}},\ \href {https://www.proquest.com/docview/3139097913} {Ph.D. thesis},\ \bibinfo  {school} {Princeton University} (\bibinfo {year} {2024})\BibitemShut {NoStop}%
\bibitem [{\citenamefont {Kawakami}\ \emph {et~al.}(2026)\citenamefont {Kawakami}, \citenamefont {Tateish}, \citenamefont {Yoshida}, \citenamefont {Yang}, \citenamefont {Nakatsuji}, \citenamefont {Chen}, \citenamefont {Aso}, \citenamefont {Yamada-Takamura}, \citenamefont {Oshima}, \citenamefont {Zhang} \emph {et~al.}}]{kawakami2026one}%
  \BibitemOpen
  \bibfield  {author} {\bibinfo {author} {\bibfnamefont {T.}~\bibnamefont {Kawakami}}, \bibinfo {author} {\bibfnamefont {H.}~\bibnamefont {Tateish}}, \bibinfo {author} {\bibfnamefont {D.}~\bibnamefont {Yoshida}}, \bibinfo {author} {\bibfnamefont {X.}~\bibnamefont {Yang}}, \bibinfo {author} {\bibfnamefont {N.}~\bibnamefont {Nakatsuji}}, \bibinfo {author} {\bibfnamefont {L.}~\bibnamefont {Chen}}, \bibinfo {author} {\bibfnamefont {K.}~\bibnamefont {Aso}}, \bibinfo {author} {\bibfnamefont {Y.}~\bibnamefont {Yamada-Takamura}}, \bibinfo {author} {\bibfnamefont {Y.}~\bibnamefont {Oshima}}, \bibinfo {author} {\bibfnamefont {Y.}~\bibnamefont {Zhang}}, \emph {et~al.},\ }\bibfield  {title} {\bibinfo {title} {{One-Dimensional Electronic States in a Moire Superlattice of Twisted Bilayer WTe2}},\ }\bibfield  {journal} {\bibinfo  {journal} {arXiv}\ }\href {https://doi.org/10.48550/arXiv.2601.21228} {10.48550/arXiv.2601.21228} (\bibinfo {year} {2026})\BibitemShut {NoStop}%
\bibitem [{\citenamefont {Liu}\ \emph {et~al.}(2025)\citenamefont {Liu}, \citenamefont {Zhang},\ and\ \citenamefont {Lu}}]{liu2025moire}%
  \BibitemOpen
  \bibfield  {author} {\bibinfo {author} {\bibfnamefont {J.}~\bibnamefont {Liu}}, \bibinfo {author} {\bibfnamefont {X.}~\bibnamefont {Zhang}},\ and\ \bibinfo {author} {\bibfnamefont {G.}~\bibnamefont {Lu}},\ }\bibfield  {title} {\bibinfo {title} {{Moire magnetism and moire excitons in twisted CrSBr bilayers}},\ }\href {https://doi.org/10.1073/pnas.2413326121} {\bibfield  {journal} {\bibinfo  {journal} {Proc. Natl. Acad. Sci. U.S.A.}\ }\textbf {\bibinfo {volume} {122}},\ \bibinfo {pages} {e2413326121} (\bibinfo {year} {2025})}\BibitemShut {NoStop}%
\bibitem [{\citenamefont {Li}\ \emph {et~al.}(2025{\natexlab{a}})\citenamefont {Li}, \citenamefont {Shubnic}, \citenamefont {Agarwal}, \citenamefont {Alfrey}, \citenamefont {Liu}, \citenamefont {Zhai}, \citenamefont {Lobanov}, \citenamefont {Uzdin}, \citenamefont {Li}, \citenamefont {Yang} \emph {et~al.}}]{li2025magneto}%
  \BibitemOpen
  \bibfield  {author} {\bibinfo {author} {\bibfnamefont {Q.}~\bibnamefont {Li}}, \bibinfo {author} {\bibfnamefont {A.}~\bibnamefont {Shubnic}}, \bibinfo {author} {\bibfnamefont {N.}~\bibnamefont {Agarwal}}, \bibinfo {author} {\bibfnamefont {A.}~\bibnamefont {Alfrey}}, \bibinfo {author} {\bibfnamefont {W.}~\bibnamefont {Liu}}, \bibinfo {author} {\bibfnamefont {Z.}~\bibnamefont {Zhai}}, \bibinfo {author} {\bibfnamefont {I.}~\bibnamefont {Lobanov}}, \bibinfo {author} {\bibfnamefont {V.}~\bibnamefont {Uzdin}}, \bibinfo {author} {\bibfnamefont {S.}~\bibnamefont {Li}}, \bibinfo {author} {\bibfnamefont {Y.}~\bibnamefont {Yang}}, \emph {et~al.},\ }\bibfield  {title} {\bibinfo {title} {{Magneto-Moire Excitons in Twisted Bilayer CrSBr}},\ }\bibfield  {journal} {\bibinfo  {journal} {arXiv}\ }\href {https://doi.org/10.48550/arXiv.2512.20507} {10.48550/arXiv.2512.20507} (\bibinfo {year} {2025}{\natexlab{a}})\BibitemShut {NoStop}%
\bibitem [{\citenamefont {Kennes}\ \emph {et~al.}(2020)\citenamefont {Kennes}, \citenamefont {Xian}, \citenamefont {Claassen},\ and\ \citenamefont {Rubio}}]{kennes2020one}%
  \BibitemOpen
  \bibfield  {author} {\bibinfo {author} {\bibfnamefont {D.~M.}\ \bibnamefont {Kennes}}, \bibinfo {author} {\bibfnamefont {L.}~\bibnamefont {Xian}}, \bibinfo {author} {\bibfnamefont {M.}~\bibnamefont {Claassen}},\ and\ \bibinfo {author} {\bibfnamefont {A.}~\bibnamefont {Rubio}},\ }\bibfield  {title} {\bibinfo {title} {{One-dimensional flat bands in twisted bilayer germanium selenide}},\ }\href {https://doi.org/10.1038/s41467-020-14947-0} {\bibfield  {journal} {\bibinfo  {journal} {Nat. Commun.}\ }\textbf {\bibinfo {volume} {11}},\ \bibinfo {pages} {1124} (\bibinfo {year} {2020})}\BibitemShut {NoStop}%
\bibitem [{\citenamefont {Zhao}\ \emph {et~al.}(2021)\citenamefont {Zhao}, \citenamefont {Wang}, \citenamefont {Uzer}, \citenamefont {Guo}, \citenamefont {Qi}, \citenamefont {Tan}, \citenamefont {Watanabe}, \citenamefont {Taniguchi}, \citenamefont {Nilges}, \citenamefont {Gao} \emph {et~al.}}]{zhao2021anisotropic}%
  \BibitemOpen
  \bibfield  {author} {\bibinfo {author} {\bibfnamefont {S.}~\bibnamefont {Zhao}}, \bibinfo {author} {\bibfnamefont {E.}~\bibnamefont {Wang}}, \bibinfo {author} {\bibfnamefont {E.~A.}\ \bibnamefont {Uzer}}, \bibinfo {author} {\bibfnamefont {S.}~\bibnamefont {Guo}}, \bibinfo {author} {\bibfnamefont {R.}~\bibnamefont {Qi}}, \bibinfo {author} {\bibfnamefont {J.}~\bibnamefont {Tan}}, \bibinfo {author} {\bibfnamefont {K.}~\bibnamefont {Watanabe}}, \bibinfo {author} {\bibfnamefont {T.}~\bibnamefont {Taniguchi}}, \bibinfo {author} {\bibfnamefont {T.}~\bibnamefont {Nilges}}, \bibinfo {author} {\bibfnamefont {P.}~\bibnamefont {Gao}}, \emph {et~al.},\ }\bibfield  {title} {\bibinfo {title} {{Anisotropic moire optical transitions in twisted monolayer/bilayer phosphorene heterostructures}},\ }\href {https://doi.org/10.1038/s41467-021-24272-9} {\bibfield  {journal} {\bibinfo  {journal} {Nat. Commun.}\ }\textbf {\bibinfo {volume} {12}},\ \bibinfo {pages} {3947} (\bibinfo {year} {2021})}\BibitemShut {NoStop}%
\bibitem [{\citenamefont {Soltero}\ \emph {et~al.}(2022)\citenamefont {Soltero}, \citenamefont {Guerrero-Sanchez}, \citenamefont {Mireles},\ and\ \citenamefont {Ruiz-Tijerina}}]{soltero2022moire}%
  \BibitemOpen
  \bibfield  {author} {\bibinfo {author} {\bibfnamefont {I.}~\bibnamefont {Soltero}}, \bibinfo {author} {\bibfnamefont {J.}~\bibnamefont {Guerrero-Sanchez}}, \bibinfo {author} {\bibfnamefont {F.}~\bibnamefont {Mireles}},\ and\ \bibinfo {author} {\bibfnamefont {D.~A.}\ \bibnamefont {Ruiz-Tijerina}},\ }\bibfield  {title} {\bibinfo {title} {{Moire band structures of twisted phosphorene bilayers}},\ }\href {https://doi.org/10.1103/PhysRevB.105.235421} {\bibfield  {journal} {\bibinfo  {journal} {Phys. Rev. B}\ }\textbf {\bibinfo {volume} {105}},\ \bibinfo {pages} {235421} (\bibinfo {year} {2022})}\BibitemShut {NoStop}%
\bibitem [{\citenamefont {Jiang}\ \emph {et~al.}(2026)\citenamefont {Jiang}, \citenamefont {An}, \citenamefont {Chen}, \citenamefont {Xu}, \citenamefont {Zhang}, \citenamefont {Fu}, \citenamefont {Dai}, \citenamefont {Yang}, \citenamefont {He}, \citenamefont {Wei} \emph {et~al.}}]{jiang2026twist}%
  \BibitemOpen
  \bibfield  {author} {\bibinfo {author} {\bibfnamefont {H.}~\bibnamefont {Jiang}}, \bibinfo {author} {\bibfnamefont {L.}~\bibnamefont {An}}, \bibinfo {author} {\bibfnamefont {X.}~\bibnamefont {Chen}}, \bibinfo {author} {\bibfnamefont {G.}~\bibnamefont {Xu}}, \bibinfo {author} {\bibfnamefont {Y.}~\bibnamefont {Zhang}}, \bibinfo {author} {\bibfnamefont {J.}~\bibnamefont {Fu}}, \bibinfo {author} {\bibfnamefont {X.}~\bibnamefont {Dai}}, \bibinfo {author} {\bibfnamefont {Y.}~\bibnamefont {Yang}}, \bibinfo {author} {\bibfnamefont {R.}~\bibnamefont {He}}, \bibinfo {author} {\bibfnamefont {X.}~\bibnamefont {Wei}}, \emph {et~al.},\ }\bibfield  {title} {\bibinfo {title} {{Twist-stacked black phosphorus for wide-spectral chiral photodetection}},\ }\href {https://doi.org/10.1038/s41467-026-68531-z} {\bibfield  {journal} {\bibinfo  {journal} {Nat. Commun.}\ }\textbf {\bibinfo {volume} {17}},\ \bibinfo {pages} {1824} (\bibinfo {year} {2026})}\BibitemShut {NoStop}%
\bibitem [{\citenamefont {Wang}\ \emph {et~al.}(2023)\citenamefont {Wang}, \citenamefont {Li}, \citenamefont {Zha}, \citenamefont {Yan}, \citenamefont {Liu},\ and\ \citenamefont {Tian}}]{wang2023tunable}%
  \BibitemOpen
  \bibfield  {author} {\bibinfo {author} {\bibfnamefont {S.~Y.}\ \bibnamefont {Wang}}, \bibinfo {author} {\bibfnamefont {D.~K.}\ \bibnamefont {Li}}, \bibinfo {author} {\bibfnamefont {M.~J.}\ \bibnamefont {Zha}}, \bibinfo {author} {\bibfnamefont {X.~Q.}\ \bibnamefont {Yan}}, \bibinfo {author} {\bibfnamefont {Z.}~\bibnamefont {Liu}},\ and\ \bibinfo {author} {\bibfnamefont {J.}~\bibnamefont {Tian}},\ }\bibfield  {title} {\bibinfo {title} {{Tunable optical activity in twisted anisotropic two-dimensional materials}},\ }\href {https://doi.org/10.1021/acsnano.3c06031} {\bibfield  {journal} {\bibinfo  {journal} {ACS Nano}\ }\textbf {\bibinfo {volume} {17}},\ \bibinfo {pages} {16230} (\bibinfo {year} {2023})}\BibitemShut {NoStop}%
\bibitem [{\citenamefont {Casimir}(1945)}]{casimir1945onsager}%
  \BibitemOpen
  \bibfield  {author} {\bibinfo {author} {\bibfnamefont {H.~B.~G.}\ \bibnamefont {Casimir}},\ }\bibfield  {title} {\bibinfo {title} {{On Onsager's principle of microscopic reversibility}},\ }\href {https://doi.org/10.1103/RevModPhys.17.343} {\bibfield  {journal} {\bibinfo  {journal} {Rev. Mod. Phys.}\ }\textbf {\bibinfo {volume} {17}},\ \bibinfo {pages} {343} (\bibinfo {year} {1945})}\BibitemShut {NoStop}%
\bibitem [{\citenamefont {Rogacheva}\ \emph {et~al.}(2006)\citenamefont {Rogacheva}, \citenamefont {Fedotov}, \citenamefont {Schwanecke},\ and\ \citenamefont {Zheludev}}]{rogacheva2006giant}%
  \BibitemOpen
  \bibfield  {author} {\bibinfo {author} {\bibfnamefont {A.}~\bibnamefont {Rogacheva}}, \bibinfo {author} {\bibfnamefont {V.}~\bibnamefont {Fedotov}}, \bibinfo {author} {\bibfnamefont {A.}~\bibnamefont {Schwanecke}},\ and\ \bibinfo {author} {\bibfnamefont {N.}~\bibnamefont {Zheludev}},\ }\bibfield  {title} {\bibinfo {title} {{Giant gyrotropy due to electromagnetic-field coupling in a bilayered chiral structure}},\ }\href {https://doi.org/10.1103/PhysRevLett.97.177401} {\bibfield  {journal} {\bibinfo  {journal} {Phys. Rev. Lett.}\ }\textbf {\bibinfo {volume} {97}},\ \bibinfo {pages} {177401} (\bibinfo {year} {2006})}\BibitemShut {NoStop}%
\bibitem [{\citenamefont {Plum}\ \emph {et~al.}(2007)\citenamefont {Plum}, \citenamefont {Fedotov}, \citenamefont {Schwanecke}, \citenamefont {Zheludev},\ and\ \citenamefont {Chen}}]{plum2007giant}%
  \BibitemOpen
  \bibfield  {author} {\bibinfo {author} {\bibfnamefont {E.}~\bibnamefont {Plum}}, \bibinfo {author} {\bibfnamefont {V.}~\bibnamefont {Fedotov}}, \bibinfo {author} {\bibfnamefont {A.}~\bibnamefont {Schwanecke}}, \bibinfo {author} {\bibfnamefont {N.}~\bibnamefont {Zheludev}},\ and\ \bibinfo {author} {\bibfnamefont {Y.}~\bibnamefont {Chen}},\ }\bibfield  {title} {\bibinfo {title} {{Giant optical gyrotropy due to electromagnetic coupling}},\ }\href {https://doi.org/10.1063/1.2745203} {\bibfield  {journal} {\bibinfo  {journal} {Appl. Phys. Lett.}\ }\textbf {\bibinfo {volume} {90}},\ \bibinfo {pages} {223113} (\bibinfo {year} {2007})}\BibitemShut {NoStop}%
\bibitem [{\citenamefont {Dos~Santos}\ \emph {et~al.}(2007)\citenamefont {Dos~Santos}, \citenamefont {Peres},\ and\ \citenamefont {Neto}}]{dossantos2007graphene}%
  \BibitemOpen
  \bibfield  {author} {\bibinfo {author} {\bibfnamefont {J.~M. B.~L.}\ \bibnamefont {Dos~Santos}}, \bibinfo {author} {\bibfnamefont {N.~M.~R.}\ \bibnamefont {Peres}},\ and\ \bibinfo {author} {\bibfnamefont {A.~H.~C.}\ \bibnamefont {Neto}},\ }\bibfield  {title} {\bibinfo {title} {{Graphene bilayer with a twist: electronic structure}},\ }\href {https://doi.org/10.1103/PhysRevLett.99.256802} {\bibfield  {journal} {\bibinfo  {journal} {Phys. Rev. Lett.}\ }\textbf {\bibinfo {volume} {99}},\ \bibinfo {pages} {256802} (\bibinfo {year} {2007})}\BibitemShut {NoStop}%
\bibitem [{\citenamefont {dos Santos}\ \emph {et~al.}(2012)\citenamefont {dos Santos}, \citenamefont {Peres},\ and\ \citenamefont {Castro~Neto}}]{dossantos2012continuum}%
  \BibitemOpen
  \bibfield  {author} {\bibinfo {author} {\bibfnamefont {J.~M. B.~L.}\ \bibnamefont {dos Santos}}, \bibinfo {author} {\bibfnamefont {N.~M.~R.}\ \bibnamefont {Peres}},\ and\ \bibinfo {author} {\bibfnamefont {A.~H.}\ \bibnamefont {Castro~Neto}},\ }\bibfield  {title} {\bibinfo {title} {{Continuum model of the twisted graphene bilayer}},\ }\href {https://doi.org/10.1103/PhysRevB.86.155449} {\bibfield  {journal} {\bibinfo  {journal} {Phys. Rev. B}\ }\textbf {\bibinfo {volume} {86}},\ \bibinfo {pages} {155449} (\bibinfo {year} {2012})}\BibitemShut {NoStop}%
\bibitem [{\citenamefont {Xiao}\ \emph {et~al.}(2026{\natexlab{a}})\citenamefont {Xiao}, \citenamefont {Inbar}, \citenamefont {Birkbeck}, \citenamefont {Gershon}, \citenamefont {Zamir}, \citenamefont {Vituri}, \citenamefont {Taniguchi}, \citenamefont {Watanabe}, \citenamefont {Berg},\ and\ \citenamefont {Ilani}}]{xiao2026imaging}%
  \BibitemOpen
  \bibfield  {author} {\bibinfo {author} {\bibfnamefont {J.}~\bibnamefont {Xiao}}, \bibinfo {author} {\bibfnamefont {A.}~\bibnamefont {Inbar}}, \bibinfo {author} {\bibfnamefont {J.}~\bibnamefont {Birkbeck}}, \bibinfo {author} {\bibfnamefont {N.}~\bibnamefont {Gershon}}, \bibinfo {author} {\bibfnamefont {Y.}~\bibnamefont {Zamir}}, \bibinfo {author} {\bibfnamefont {Y.}~\bibnamefont {Vituri}}, \bibinfo {author} {\bibfnamefont {T.}~\bibnamefont {Taniguchi}}, \bibinfo {author} {\bibfnamefont {K.}~\bibnamefont {Watanabe}}, \bibinfo {author} {\bibfnamefont {E.}~\bibnamefont {Berg}},\ and\ \bibinfo {author} {\bibfnamefont {S.}~\bibnamefont {Ilani}},\ }\bibfield  {title} {\bibinfo {title} {{Imaging the flat bands of magic-angle graphene reshaped by interactions}},\ }\href {https://doi.org/10.1038/s41586-026-10378-x} {\bibfield  {journal} {\bibinfo  {journal} {Nature}\ }\textbf {\bibinfo {volume} {653}},\ \bibinfo {pages} {68} (\bibinfo {year} {2026}{\natexlab{a}})}\BibitemShut {NoStop}%
\bibitem [{\citenamefont {Inbar}\ \emph {et~al.}(2023)\citenamefont {Inbar}, \citenamefont {Birkbeck}, \citenamefont {Xiao}, \citenamefont {Taniguchi}, \citenamefont {Watanabe}, \citenamefont {Yan}, \citenamefont {Oreg}, \citenamefont {Stern}, \citenamefont {Berg},\ and\ \citenamefont {Ilani}}]{inbar2023quantum}%
  \BibitemOpen
  \bibfield  {author} {\bibinfo {author} {\bibfnamefont {A.}~\bibnamefont {Inbar}}, \bibinfo {author} {\bibfnamefont {J.}~\bibnamefont {Birkbeck}}, \bibinfo {author} {\bibfnamefont {J.}~\bibnamefont {Xiao}}, \bibinfo {author} {\bibfnamefont {T.}~\bibnamefont {Taniguchi}}, \bibinfo {author} {\bibfnamefont {K.}~\bibnamefont {Watanabe}}, \bibinfo {author} {\bibfnamefont {B.}~\bibnamefont {Yan}}, \bibinfo {author} {\bibfnamefont {Y.}~\bibnamefont {Oreg}}, \bibinfo {author} {\bibfnamefont {A.}~\bibnamefont {Stern}}, \bibinfo {author} {\bibfnamefont {E.}~\bibnamefont {Berg}},\ and\ \bibinfo {author} {\bibfnamefont {S.}~\bibnamefont {Ilani}},\ }\bibfield  {title} {\bibinfo {title} {{The quantum twisting microscope}},\ }\href {https://doi.org/10.1038/s41586-022-05685-y} {\bibfield  {journal} {\bibinfo  {journal} {Nature}\ }\textbf {\bibinfo {volume} {614}},\ \bibinfo {pages} {682} (\bibinfo {year} {2023})}\BibitemShut {NoStop}%
\bibitem [{\citenamefont {Wei}\ \emph {et~al.}(2025)\citenamefont {Wei}, \citenamefont {von Oppen},\ and\ \citenamefont {Glazman}}]{wei2025dirac}%
  \BibitemOpen
  \bibfield  {author} {\bibinfo {author} {\bibfnamefont {N.}~\bibnamefont {Wei}}, \bibinfo {author} {\bibfnamefont {F.}~\bibnamefont {von Oppen}},\ and\ \bibinfo {author} {\bibfnamefont {L.~I.}\ \bibnamefont {Glazman}},\ }\bibfield  {title} {\bibinfo {title} {{Dirac-point spectroscopy of flat-band systems with the quantum twisting microscope}},\ }\href {https://doi.org/10.1103/PhysRevB.111.085128} {\bibfield  {journal} {\bibinfo  {journal} {Phys. Rev. B}\ }\textbf {\bibinfo {volume} {111}},\ \bibinfo {pages} {085128} (\bibinfo {year} {2025})}\BibitemShut {NoStop}%
\bibitem [{\citenamefont {Birkbeck}\ \emph {et~al.}(2025)\citenamefont {Birkbeck}, \citenamefont {Xiao}, \citenamefont {Inbar}, \citenamefont {Taniguchi}, \citenamefont {Watanabe}, \citenamefont {Berg}, \citenamefont {Glazman}, \citenamefont {Guinea}, \citenamefont {von Oppen},\ and\ \citenamefont {Ilani}}]{birkbeck2025quantum}%
  \BibitemOpen
  \bibfield  {author} {\bibinfo {author} {\bibfnamefont {J.}~\bibnamefont {Birkbeck}}, \bibinfo {author} {\bibfnamefont {J.}~\bibnamefont {Xiao}}, \bibinfo {author} {\bibfnamefont {A.}~\bibnamefont {Inbar}}, \bibinfo {author} {\bibfnamefont {T.}~\bibnamefont {Taniguchi}}, \bibinfo {author} {\bibfnamefont {K.}~\bibnamefont {Watanabe}}, \bibinfo {author} {\bibfnamefont {E.}~\bibnamefont {Berg}}, \bibinfo {author} {\bibfnamefont {L.}~\bibnamefont {Glazman}}, \bibinfo {author} {\bibfnamefont {F.}~\bibnamefont {Guinea}}, \bibinfo {author} {\bibfnamefont {F.}~\bibnamefont {von Oppen}},\ and\ \bibinfo {author} {\bibfnamefont {S.}~\bibnamefont {Ilani}},\ }\bibfield  {title} {\bibinfo {title} {{Quantum twisting microscopy of phonons in twisted bilayer graphene}},\ }\href {https://doi.org/10.1038/s41586-025-08881-8} {\bibfield  {journal} {\bibinfo  {journal} {Nature}\ }\textbf {\bibinfo {volume} {641}},\ \bibinfo {pages} {345} (\bibinfo {year} {2025})}\BibitemShut {NoStop}%
\bibitem [{\citenamefont {Lee}\ \emph {et~al.}(2026)\citenamefont {Lee}, \citenamefont {Das}, \citenamefont {Herzog-Arbeitman}, \citenamefont {Papp}, \citenamefont {Li}, \citenamefont {Daschner}, \citenamefont {Zhou}, \citenamefont {Bhatt}, \citenamefont {Currle}, \citenamefont {Yu} \emph {et~al.}}]{lee2026revealing}%
  \BibitemOpen
  \bibfield  {author} {\bibinfo {author} {\bibfnamefont {M.}~\bibnamefont {Lee}}, \bibinfo {author} {\bibfnamefont {I.}~\bibnamefont {Das}}, \bibinfo {author} {\bibfnamefont {J.}~\bibnamefont {Herzog-Arbeitman}}, \bibinfo {author} {\bibfnamefont {J.}~\bibnamefont {Papp}}, \bibinfo {author} {\bibfnamefont {J.}~\bibnamefont {Li}}, \bibinfo {author} {\bibfnamefont {M.}~\bibnamefont {Daschner}}, \bibinfo {author} {\bibfnamefont {Z.}~\bibnamefont {Zhou}}, \bibinfo {author} {\bibfnamefont {M.}~\bibnamefont {Bhatt}}, \bibinfo {author} {\bibfnamefont {M.}~\bibnamefont {Currle}}, \bibinfo {author} {\bibfnamefont {J.}~\bibnamefont {Yu}}, \emph {et~al.},\ }\bibfield  {title} {\bibinfo {title} {{Revealing Electron--Electron Interactions in Graphene at Room Temperature with a Quantum Twisting Microscope}},\ }\href {https://doi.org/10.1021/acs.nanolett.5c05015} {\bibfield  {journal} {\bibinfo  {journal} {Nano Lett.}\ }\textbf {\bibinfo {volume} {26}},\ \bibinfo {pages} {4046} (\bibinfo {year} {2026})}\BibitemShut {NoStop}%
\bibitem [{\citenamefont {Nam}\ and\ \citenamefont {Koshino}(2017)}]{nam2017lattice}%
  \BibitemOpen
  \bibfield  {author} {\bibinfo {author} {\bibfnamefont {N.~N.}\ \bibnamefont {Nam}}\ and\ \bibinfo {author} {\bibfnamefont {M.}~\bibnamefont {Koshino}},\ }\bibfield  {title} {\bibinfo {title} {{Lattice relaxation and energy band modulation in twisted bilayer graphenes}},\ }\href {https://doi.org/10.1103/PhysRevB.96.075311} {\bibfield  {journal} {\bibinfo  {journal} {Phys. Rev. B}\ }\textbf {\bibinfo {volume} {96}},\ \bibinfo {pages} {075311} (\bibinfo {year} {2017})}\BibitemShut {NoStop}%
\bibitem [{\citenamefont {Ezzi}\ \emph {et~al.}(2024)\citenamefont {Ezzi}, \citenamefont {Pallewela}, \citenamefont {De~Beule}, \citenamefont {Mele},\ and\ \citenamefont {Adam}}]{ezzi2024analytical}%
  \BibitemOpen
  \bibfield  {author} {\bibinfo {author} {\bibfnamefont {M.~M.~A.}\ \bibnamefont {Ezzi}}, \bibinfo {author} {\bibfnamefont {G.~N.}\ \bibnamefont {Pallewela}}, \bibinfo {author} {\bibfnamefont {C.}~\bibnamefont {De~Beule}}, \bibinfo {author} {\bibfnamefont {E.~J.}\ \bibnamefont {Mele}},\ and\ \bibinfo {author} {\bibfnamefont {S.}~\bibnamefont {Adam}},\ }\bibfield  {title} {\bibinfo {title} {{Analytical model for atomic relaxation in twisted moir{\'e} materials}},\ }\href {https://doi.org/10.1103/PhysRevLett.133.266201} {\bibfield  {journal} {\bibinfo  {journal} {Physical Review Letters}\ }\textbf {\bibinfo {volume} {133}},\ \bibinfo {pages} {266201} (\bibinfo {year} {2024})}\BibitemShut {NoStop}%
\bibitem [{\citenamefont {Mahan}(2000)}]{mahan2000many}%
  \BibitemOpen
  \bibfield  {author} {\bibinfo {author} {\bibfnamefont {G.~D.}\ \bibnamefont {Mahan}},\ }\href {https://doi.org/10.1007/978-1-4757-5714-9} {\emph {\bibinfo {title} {{Many-Particle Physics}}}}\ (\bibinfo  {publisher} {Springer},\ \bibinfo {year} {2000})\BibitemShut {NoStop}%
\bibitem [{\citenamefont {Wu}\ \emph {et~al.}(2018)\citenamefont {Wu}, \citenamefont {Lovorn}, \citenamefont {Tutuc},\ and\ \citenamefont {MacDonald}}]{wu2018hubbard}%
  \BibitemOpen
  \bibfield  {author} {\bibinfo {author} {\bibfnamefont {F.}~\bibnamefont {Wu}}, \bibinfo {author} {\bibfnamefont {T.}~\bibnamefont {Lovorn}}, \bibinfo {author} {\bibfnamefont {E.}~\bibnamefont {Tutuc}},\ and\ \bibinfo {author} {\bibfnamefont {A.~H.}\ \bibnamefont {MacDonald}},\ }\bibfield  {title} {\bibinfo {title} {{Hubbard model physics in transition metal dichalcogenide moire bands}},\ }\href {https://doi.org/10.1103/PhysRevLett.121.026402} {\bibfield  {journal} {\bibinfo  {journal} {Phys. Rev. Lett.}\ }\textbf {\bibinfo {volume} {121}},\ \bibinfo {pages} {026402} (\bibinfo {year} {2018})}\BibitemShut {NoStop}%
\bibitem [{\citenamefont {Jia}\ \emph {et~al.}(2024)\citenamefont {Jia}, \citenamefont {Yu}, \citenamefont {Liu}, \citenamefont {Herzog-Arbeitman}, \citenamefont {Qi}, \citenamefont {Pi}, \citenamefont {Regnault}, \citenamefont {Weng}, \citenamefont {Bernevig},\ and\ \citenamefont {Wu}}]{jia2024moire}%
  \BibitemOpen
  \bibfield  {author} {\bibinfo {author} {\bibfnamefont {Y.}~\bibnamefont {Jia}}, \bibinfo {author} {\bibfnamefont {J.}~\bibnamefont {Yu}}, \bibinfo {author} {\bibfnamefont {J.}~\bibnamefont {Liu}}, \bibinfo {author} {\bibfnamefont {J.}~\bibnamefont {Herzog-Arbeitman}}, \bibinfo {author} {\bibfnamefont {Z.}~\bibnamefont {Qi}}, \bibinfo {author} {\bibfnamefont {H.}~\bibnamefont {Pi}}, \bibinfo {author} {\bibfnamefont {N.}~\bibnamefont {Regnault}}, \bibinfo {author} {\bibfnamefont {H.}~\bibnamefont {Weng}}, \bibinfo {author} {\bibfnamefont {B.~A.}\ \bibnamefont {Bernevig}},\ and\ \bibinfo {author} {\bibfnamefont {Q.}~\bibnamefont {Wu}},\ }\bibfield  {title} {\bibinfo {title} {{Moire fractional Chern insulators. I. First-principles calculations and continuum models of twisted bilayer MoTe2}},\ }\href {https://doi.org/10.1103/PhysRevB.109.205121} {\bibfield  {journal} {\bibinfo  {journal} {Phys. Rev. B}\ }\textbf {\bibinfo {volume} {109}},\ \bibinfo {pages} {205121} (\bibinfo {year} {2024})}\BibitemShut {NoStop}%
\bibitem [{\citenamefont {Wu}\ \emph {et~al.}(2019)\citenamefont {Wu}, \citenamefont {Lovorn}, \citenamefont {Tutuc}, \citenamefont {Martin},\ and\ \citenamefont {MacDonald}}]{wu2019topological}%
  \BibitemOpen
  \bibfield  {author} {\bibinfo {author} {\bibfnamefont {F.}~\bibnamefont {Wu}}, \bibinfo {author} {\bibfnamefont {T.}~\bibnamefont {Lovorn}}, \bibinfo {author} {\bibfnamefont {E.}~\bibnamefont {Tutuc}}, \bibinfo {author} {\bibfnamefont {I.}~\bibnamefont {Martin}},\ and\ \bibinfo {author} {\bibfnamefont {A.}~\bibnamefont {MacDonald}},\ }\bibfield  {title} {\bibinfo {title} {{Topological insulators in twisted transition metal dichalcogenide homobilayers}},\ }\href {https://doi.org/10.1103/PhysRevLett.122.086402} {\bibfield  {journal} {\bibinfo  {journal} {Phys. Rev. Lett.}\ }\textbf {\bibinfo {volume} {122}},\ \bibinfo {pages} {086402} (\bibinfo {year} {2019})}\BibitemShut {NoStop}%
\bibitem [{\citenamefont {Xu}\ \emph {et~al.}(2014)\citenamefont {Xu}, \citenamefont {Yao}, \citenamefont {Xiao},\ and\ \citenamefont {Heinz}}]{xu2014spin}%
  \BibitemOpen
  \bibfield  {author} {\bibinfo {author} {\bibfnamefont {X.}~\bibnamefont {Xu}}, \bibinfo {author} {\bibfnamefont {W.}~\bibnamefont {Yao}}, \bibinfo {author} {\bibfnamefont {D.}~\bibnamefont {Xiao}},\ and\ \bibinfo {author} {\bibfnamefont {T.~F.}\ \bibnamefont {Heinz}},\ }\bibfield  {title} {\bibinfo {title} {{Spin and pseudospins in layered transition metal dichalcogenides}},\ }\href {https://doi.org/10.1038/nphys2942} {\bibfield  {journal} {\bibinfo  {journal} {Nat. Phys.}\ }\textbf {\bibinfo {volume} {10}},\ \bibinfo {pages} {343} (\bibinfo {year} {2014})}\BibitemShut {NoStop}%
\bibitem [{\citenamefont {Li}\ \emph {et~al.}(2025{\natexlab{b}})\citenamefont {Li}, \citenamefont {Redekop}, \citenamefont {Wang~Beach}, \citenamefont {Zhang}, \citenamefont {Zhang}, \citenamefont {Liu}, \citenamefont {Holtzmann}, \citenamefont {Hu}, \citenamefont {Anderson}, \citenamefont {Park} \emph {et~al.}}]{li2025universal}%
  \BibitemOpen
  \bibfield  {author} {\bibinfo {author} {\bibfnamefont {W.}~\bibnamefont {Li}}, \bibinfo {author} {\bibfnamefont {E.}~\bibnamefont {Redekop}}, \bibinfo {author} {\bibfnamefont {C.}~\bibnamefont {Wang~Beach}}, \bibinfo {author} {\bibfnamefont {C.}~\bibnamefont {Zhang}}, \bibinfo {author} {\bibfnamefont {X.}~\bibnamefont {Zhang}}, \bibinfo {author} {\bibfnamefont {X.}~\bibnamefont {Liu}}, \bibinfo {author} {\bibfnamefont {W.}~\bibnamefont {Holtzmann}}, \bibinfo {author} {\bibfnamefont {C.}~\bibnamefont {Hu}}, \bibinfo {author} {\bibfnamefont {E.}~\bibnamefont {Anderson}}, \bibinfo {author} {\bibfnamefont {H.}~\bibnamefont {Park}}, \emph {et~al.},\ }\bibfield  {title} {\bibinfo {title} {{Universal magnetic phases in twisted bilayer MoTe2}},\ }\href {https://doi.org/10.1021/acs.nanolett.5c04751} {\bibfield  {journal} {\bibinfo  {journal} {Nano Lett.}\ }\textbf {\bibinfo {volume} {25}},\ \bibinfo {pages} {18044} (\bibinfo {year} {2025}{\natexlab{b}})}\BibitemShut {NoStop}%
\bibitem [{\citenamefont {Zeng}\ \emph {et~al.}(2023)\citenamefont {Zeng}, \citenamefont {Xia}, \citenamefont {Kang}, \citenamefont {Zhu}, \citenamefont {Kn{\"u}ppel}, \citenamefont {Vaswani}, \citenamefont {Watanabe}, \citenamefont {Taniguchi}, \citenamefont {Mak},\ and\ \citenamefont {Shan}}]{zeng2023integer}%
  \BibitemOpen
  \bibfield  {author} {\bibinfo {author} {\bibfnamefont {Y.}~\bibnamefont {Zeng}}, \bibinfo {author} {\bibfnamefont {Z.}~\bibnamefont {Xia}}, \bibinfo {author} {\bibfnamefont {K.}~\bibnamefont {Kang}}, \bibinfo {author} {\bibfnamefont {J.}~\bibnamefont {Zhu}}, \bibinfo {author} {\bibfnamefont {P.}~\bibnamefont {Kn{\"u}ppel}}, \bibinfo {author} {\bibfnamefont {C.}~\bibnamefont {Vaswani}}, \bibinfo {author} {\bibfnamefont {K.}~\bibnamefont {Watanabe}}, \bibinfo {author} {\bibfnamefont {T.}~\bibnamefont {Taniguchi}}, \bibinfo {author} {\bibfnamefont {K.~F.}\ \bibnamefont {Mak}},\ and\ \bibinfo {author} {\bibfnamefont {J.}~\bibnamefont {Shan}},\ }\bibfield  {title} {\bibinfo {title} {{Integer and fractional Chern insulators in twisted bilayer MoTe2}},\ }\href {https://doi.org/10.1038/s41586-023-06452-3} {\bibfield  {journal} {\bibinfo  {journal} {Nature}\ }\textbf {\bibinfo {volume} {622}},\ \bibinfo {pages} {69} (\bibinfo {year} {2023})}\BibitemShut {NoStop}%
\bibitem [{\citenamefont {Kang}\ \emph {et~al.}(2024)\citenamefont {Kang}, \citenamefont {Shen}, \citenamefont {Qiu}, \citenamefont {Zeng}, \citenamefont {Xia}, \citenamefont {Watanabe}, \citenamefont {Taniguchi}, \citenamefont {Shan},\ and\ \citenamefont {Mak}}]{kang2024evidence}%
  \BibitemOpen
  \bibfield  {author} {\bibinfo {author} {\bibfnamefont {K.}~\bibnamefont {Kang}}, \bibinfo {author} {\bibfnamefont {B.}~\bibnamefont {Shen}}, \bibinfo {author} {\bibfnamefont {Y.}~\bibnamefont {Qiu}}, \bibinfo {author} {\bibfnamefont {Y.}~\bibnamefont {Zeng}}, \bibinfo {author} {\bibfnamefont {Z.}~\bibnamefont {Xia}}, \bibinfo {author} {\bibfnamefont {K.}~\bibnamefont {Watanabe}}, \bibinfo {author} {\bibfnamefont {T.}~\bibnamefont {Taniguchi}}, \bibinfo {author} {\bibfnamefont {J.}~\bibnamefont {Shan}},\ and\ \bibinfo {author} {\bibfnamefont {K.~F.}\ \bibnamefont {Mak}},\ }\bibfield  {title} {\bibinfo {title} {{Evidence of the fractional quantum spin Hall effect in moire MoTe2}},\ }\href {https://doi.org/10.1038/s41586-024-07214-5} {\bibfield  {journal} {\bibinfo  {journal} {Nature}\ }\textbf {\bibinfo {volume} {628}},\ \bibinfo {pages} {522} (\bibinfo {year} {2024})}\BibitemShut {NoStop}%
\bibitem [{\citenamefont {Wu}\ \emph {et~al.}(2026)\citenamefont {Wu}, \citenamefont {Li}, \citenamefont {Ouyang}, \citenamefont {Jiang}, \citenamefont {Qiu}, \citenamefont {Zhang}, \citenamefont {Huo}, \citenamefont {Yang}, \citenamefont {Tian}, \citenamefont {Wan} \emph {et~al.}}]{wu2026observation}%
  \BibitemOpen
  \bibfield  {author} {\bibinfo {author} {\bibfnamefont {M.}~\bibnamefont {Wu}}, \bibinfo {author} {\bibfnamefont {L.}~\bibnamefont {Li}}, \bibinfo {author} {\bibfnamefont {Y.}~\bibnamefont {Ouyang}}, \bibinfo {author} {\bibfnamefont {Y.}~\bibnamefont {Jiang}}, \bibinfo {author} {\bibfnamefont {W.}~\bibnamefont {Qiu}}, \bibinfo {author} {\bibfnamefont {Z.}~\bibnamefont {Zhang}}, \bibinfo {author} {\bibfnamefont {Z.}~\bibnamefont {Huo}}, \bibinfo {author} {\bibfnamefont {Q.}~\bibnamefont {Yang}}, \bibinfo {author} {\bibfnamefont {M.}~\bibnamefont {Tian}}, \bibinfo {author} {\bibfnamefont {N.}~\bibnamefont {Wan}}, \emph {et~al.},\ }\bibfield  {title} {\bibinfo {title} {{Observation of a Reconstructed Chern Insulator in Twisted Bilayer MoTe2}},\ }\bibfield  {journal} {\bibinfo  {journal} {arXiv}\ }\href {https://doi.org/10.48550/arXiv.2603.16374} {10.48550/arXiv.2603.16374} (\bibinfo {year} {2026})\BibitemShut {NoStop}%
\bibitem [{\citenamefont {Cai}\ \emph {et~al.}(2023)\citenamefont {Cai}, \citenamefont {Anderson}, \citenamefont {Wang}, \citenamefont {Zhang}, \citenamefont {Liu}, \citenamefont {Holtzmann}, \citenamefont {Zhang}, \citenamefont {Fan}, \citenamefont {Taniguchi}, \citenamefont {Watanabe} \emph {et~al.}}]{cai2023signatures}%
  \BibitemOpen
  \bibfield  {author} {\bibinfo {author} {\bibfnamefont {J.}~\bibnamefont {Cai}}, \bibinfo {author} {\bibfnamefont {E.}~\bibnamefont {Anderson}}, \bibinfo {author} {\bibfnamefont {C.}~\bibnamefont {Wang}}, \bibinfo {author} {\bibfnamefont {X.}~\bibnamefont {Zhang}}, \bibinfo {author} {\bibfnamefont {X.}~\bibnamefont {Liu}}, \bibinfo {author} {\bibfnamefont {W.}~\bibnamefont {Holtzmann}}, \bibinfo {author} {\bibfnamefont {Y.}~\bibnamefont {Zhang}}, \bibinfo {author} {\bibfnamefont {F.}~\bibnamefont {Fan}}, \bibinfo {author} {\bibfnamefont {T.}~\bibnamefont {Taniguchi}}, \bibinfo {author} {\bibfnamefont {K.}~\bibnamefont {Watanabe}}, \emph {et~al.},\ }\bibfield  {title} {\bibinfo {title} {{Signatures of fractional quantum anomalous Hall states in twisted MoTe2}},\ }\href {https://doi.org/10.1038/s41586-023-06289-w} {\bibfield  {journal} {\bibinfo  {journal} {Nature}\ }\textbf {\bibinfo {volume} {622}},\ \bibinfo {pages} {63} (\bibinfo {year} {2023})}\BibitemShut {NoStop}%
\bibitem [{\citenamefont {Park}\ \emph {et~al.}(2023)\citenamefont {Park}, \citenamefont {Cai}, \citenamefont {Anderson}, \citenamefont {Zhang}, \citenamefont {Zhu}, \citenamefont {Liu}, \citenamefont {Wang}, \citenamefont {Holtzmann}, \citenamefont {Hu}, \citenamefont {Liu} \emph {et~al.}}]{park2023observation}%
  \BibitemOpen
  \bibfield  {author} {\bibinfo {author} {\bibfnamefont {H.}~\bibnamefont {Park}}, \bibinfo {author} {\bibfnamefont {J.}~\bibnamefont {Cai}}, \bibinfo {author} {\bibfnamefont {E.}~\bibnamefont {Anderson}}, \bibinfo {author} {\bibfnamefont {Y.}~\bibnamefont {Zhang}}, \bibinfo {author} {\bibfnamefont {J.}~\bibnamefont {Zhu}}, \bibinfo {author} {\bibfnamefont {X.}~\bibnamefont {Liu}}, \bibinfo {author} {\bibfnamefont {C.}~\bibnamefont {Wang}}, \bibinfo {author} {\bibfnamefont {W.}~\bibnamefont {Holtzmann}}, \bibinfo {author} {\bibfnamefont {C.}~\bibnamefont {Hu}}, \bibinfo {author} {\bibfnamefont {Z.}~\bibnamefont {Liu}}, \emph {et~al.},\ }\bibfield  {title} {\bibinfo {title} {{Observation of fractionally quantized anomalous Hall effect}},\ }\href {https://doi.org/10.1038/s41586-023-06536-0} {\bibfield  {journal} {\bibinfo  {journal} {Nature}\ }\textbf {\bibinfo {volume} {622}},\ \bibinfo {pages} {74} (\bibinfo {year} {2023})}\BibitemShut {NoStop}%
\bibitem [{\citenamefont {Dai}\ \emph {et~al.}(2021)\citenamefont {Dai}, \citenamefont {He},\ and\ \citenamefont {Li}}]{dai2021effects}%
  \BibitemOpen
  \bibfield  {author} {\bibinfo {author} {\bibfnamefont {Z.~B.}\ \bibnamefont {Dai}}, \bibinfo {author} {\bibfnamefont {Y.}~\bibnamefont {He}},\ and\ \bibinfo {author} {\bibfnamefont {Z.}~\bibnamefont {Li}},\ }\bibfield  {title} {\bibinfo {title} {{Effects of heterostrain and lattice relaxation on the optical conductivity of twisted bilayer graphene}},\ }\href {https://doi.org/10.1103/PhysRevB.104.045403} {\bibfield  {journal} {\bibinfo  {journal} {Phys. Rev. B}\ }\textbf {\bibinfo {volume} {104}},\ \bibinfo {pages} {045403} (\bibinfo {year} {2021})}\BibitemShut {NoStop}%
\bibitem [{\citenamefont {Bi}\ \emph {et~al.}(2019)\citenamefont {Bi}, \citenamefont {Yuan},\ and\ \citenamefont {Fu}}]{bi2019designing}%
  \BibitemOpen
  \bibfield  {author} {\bibinfo {author} {\bibfnamefont {Z.}~\bibnamefont {Bi}}, \bibinfo {author} {\bibfnamefont {N.~F.}\ \bibnamefont {Yuan}},\ and\ \bibinfo {author} {\bibfnamefont {L.}~\bibnamefont {Fu}},\ }\bibfield  {title} {\bibinfo {title} {{Designing flat bands by strain}},\ }\href {https://doi.org/10.1103/PhysRevB.100.035448} {\bibfield  {journal} {\bibinfo  {journal} {Phys. Rev. B}\ }\textbf {\bibinfo {volume} {100}},\ \bibinfo {pages} {035448} (\bibinfo {year} {2019})}\BibitemShut {NoStop}%
\bibitem [{\citenamefont {Escudero}\ \emph {et~al.}(2024)\citenamefont {Escudero}, \citenamefont {Sinner}, \citenamefont {Zhan}, \citenamefont {Pantaleon},\ and\ \citenamefont {Guinea}}]{escudero2024designing}%
  \BibitemOpen
  \bibfield  {author} {\bibinfo {author} {\bibfnamefont {F.}~\bibnamefont {Escudero}}, \bibinfo {author} {\bibfnamefont {A.}~\bibnamefont {Sinner}}, \bibinfo {author} {\bibfnamefont {Z.}~\bibnamefont {Zhan}}, \bibinfo {author} {\bibfnamefont {P.~A.}\ \bibnamefont {Pantaleon}},\ and\ \bibinfo {author} {\bibfnamefont {F.}~\bibnamefont {Guinea}},\ }\bibfield  {title} {\bibinfo {title} {{Designing moire patterns by strain}},\ }\href {https://doi.org/10.1103/PhysRevResearch.6.023203} {\bibfield  {journal} {\bibinfo  {journal} {Phys. Rev. Res.}\ }\textbf {\bibinfo {volume} {6}},\ \bibinfo {pages} {023203} (\bibinfo {year} {2024})}\BibitemShut {NoStop}%
\bibitem [{\citenamefont {Mortazavi}\ \emph {et~al.}(2018)\citenamefont {Mortazavi}, \citenamefont {Berdiyorov}, \citenamefont {Makaremi},\ and\ \citenamefont {Rabczuk}}]{mortazavi2018mechanical}%
  \BibitemOpen
  \bibfield  {author} {\bibinfo {author} {\bibfnamefont {B.}~\bibnamefont {Mortazavi}}, \bibinfo {author} {\bibfnamefont {G.~R.}\ \bibnamefont {Berdiyorov}}, \bibinfo {author} {\bibfnamefont {M.}~\bibnamefont {Makaremi}},\ and\ \bibinfo {author} {\bibfnamefont {T.}~\bibnamefont {Rabczuk}},\ }\bibfield  {title} {\bibinfo {title} {{Mechanical responses of two-dimensional MoTe2; pristine 2H, 1T and 1T' and 1T'/2H heterostructure}},\ }\href {https://doi.org/10.1016/j.eml.2018.01.005} {\bibfield  {journal} {\bibinfo  {journal} {Extreme Mech. Lett.}\ }\textbf {\bibinfo {volume} {20}},\ \bibinfo {pages} {65} (\bibinfo {year} {2018})}\BibitemShut {NoStop}%
\bibitem [{\citenamefont {Woo}\ \emph {et~al.}(2016)\citenamefont {Woo}, \citenamefont {Park},\ and\ \citenamefont {Son}}]{woo2016poisson}%
  \BibitemOpen
  \bibfield  {author} {\bibinfo {author} {\bibfnamefont {S.}~\bibnamefont {Woo}}, \bibinfo {author} {\bibfnamefont {H.~C.}\ \bibnamefont {Park}},\ and\ \bibinfo {author} {\bibfnamefont {Y.~W.}\ \bibnamefont {Son}},\ }\bibfield  {title} {\bibinfo {title} {{Poisson's ratio in layered two-dimensional crystals}},\ }\href {https://doi.org/10.1103/PhysRevB.93.075420} {\bibfield  {journal} {\bibinfo  {journal} {Phys. Rev. B}\ }\textbf {\bibinfo {volume} {93}},\ \bibinfo {pages} {075420} (\bibinfo {year} {2016})}\BibitemShut {NoStop}%
\bibitem [{\citenamefont {Thompson}\ \emph {et~al.}(2025)\citenamefont {Thompson}, \citenamefont {Chu}, \citenamefont {Mesple}, \citenamefont {Zhang}, \citenamefont {Hu}, \citenamefont {Zhao}, \citenamefont {Park}, \citenamefont {Cai}, \citenamefont {Anderson}, \citenamefont {Watanabe} \emph {et~al.}}]{thompson2025microscopic}%
  \BibitemOpen
  \bibfield  {author} {\bibinfo {author} {\bibfnamefont {E.}~\bibnamefont {Thompson}}, \bibinfo {author} {\bibfnamefont {K.~T.}\ \bibnamefont {Chu}}, \bibinfo {author} {\bibfnamefont {F.}~\bibnamefont {Mesple}}, \bibinfo {author} {\bibfnamefont {X.-W.}\ \bibnamefont {Zhang}}, \bibinfo {author} {\bibfnamefont {C.}~\bibnamefont {Hu}}, \bibinfo {author} {\bibfnamefont {Y.}~\bibnamefont {Zhao}}, \bibinfo {author} {\bibfnamefont {H.}~\bibnamefont {Park}}, \bibinfo {author} {\bibfnamefont {J.}~\bibnamefont {Cai}}, \bibinfo {author} {\bibfnamefont {E.}~\bibnamefont {Anderson}}, \bibinfo {author} {\bibfnamefont {K.}~\bibnamefont {Watanabe}}, \emph {et~al.},\ }\bibfield  {title} {\bibinfo {title} {{Microscopic signatures of topology in twisted MoTe2}},\ }\href {https://doi.org/10.1038/s41567-025-02877-x} {\bibfield  {journal} {\bibinfo  {journal} {Nat. Phys.}\ }\textbf {\bibinfo {volume} {21}},\ \bibinfo {pages} {1224} (\bibinfo {year} {2025})}\BibitemShut {NoStop}%
\bibitem [{\citenamefont {Lu}\ \emph {et~al.}(2024)\citenamefont {Lu}, \citenamefont {Han}, \citenamefont {Yao}, \citenamefont {Reddy}, \citenamefont {Yang}, \citenamefont {Seo}, \citenamefont {Watanabe}, \citenamefont {Taniguchi}, \citenamefont {Fu},\ and\ \citenamefont {Ju}}]{lu2024fractional}%
  \BibitemOpen
  \bibfield  {author} {\bibinfo {author} {\bibfnamefont {Z.}~\bibnamefont {Lu}}, \bibinfo {author} {\bibfnamefont {T.}~\bibnamefont {Han}}, \bibinfo {author} {\bibfnamefont {Y.}~\bibnamefont {Yao}}, \bibinfo {author} {\bibfnamefont {A.~P.}\ \bibnamefont {Reddy}}, \bibinfo {author} {\bibfnamefont {J.}~\bibnamefont {Yang}}, \bibinfo {author} {\bibfnamefont {J.}~\bibnamefont {Seo}}, \bibinfo {author} {\bibfnamefont {K.}~\bibnamefont {Watanabe}}, \bibinfo {author} {\bibfnamefont {T.}~\bibnamefont {Taniguchi}}, \bibinfo {author} {\bibfnamefont {L.}~\bibnamefont {Fu}},\ and\ \bibinfo {author} {\bibfnamefont {L.}~\bibnamefont {Ju}},\ }\bibfield  {title} {\bibinfo {title} {{Fractional quantum anomalous Hall effect in multilayer graphene}},\ }\href {https://doi.org/10.1038/s41586-023-07010-7} {\bibfield  {journal} {\bibinfo  {journal} {Nature}\ }\textbf {\bibinfo {volume} {626}},\ \bibinfo {pages} {759} (\bibinfo {year} {2024})}\BibitemShut {NoStop}%
\bibitem [{\citenamefont {Ji}\ \emph {et~al.}(2024)\citenamefont {Ji}, \citenamefont {Zhao}, \citenamefont {Chen}, \citenamefont {Zhu}, \citenamefont {Wang}, \citenamefont {Liu}, \citenamefont {Modi}, \citenamefont {Mele}, \citenamefont {Jin},\ and\ \citenamefont {Agarwal}}]{ji2024opto}%
  \BibitemOpen
  \bibfield  {author} {\bibinfo {author} {\bibfnamefont {Z.}~\bibnamefont {Ji}}, \bibinfo {author} {\bibfnamefont {Y.}~\bibnamefont {Zhao}}, \bibinfo {author} {\bibfnamefont {Y.}~\bibnamefont {Chen}}, \bibinfo {author} {\bibfnamefont {Z.}~\bibnamefont {Zhu}}, \bibinfo {author} {\bibfnamefont {Y.}~\bibnamefont {Wang}}, \bibinfo {author} {\bibfnamefont {W.}~\bibnamefont {Liu}}, \bibinfo {author} {\bibfnamefont {G.}~\bibnamefont {Modi}}, \bibinfo {author} {\bibfnamefont {E.~J.}\ \bibnamefont {Mele}}, \bibinfo {author} {\bibfnamefont {S.}~\bibnamefont {Jin}},\ and\ \bibinfo {author} {\bibfnamefont {R.}~\bibnamefont {Agarwal}},\ }\bibfield  {title} {\bibinfo {title} {{Opto-twistronic Hall effect in a three-dimensional spiral lattice}},\ }\href {https://doi.org/10.1038/s41586-024-07949-1} {\bibfield  {journal} {\bibinfo  {journal} {Nature}\ }\textbf {\bibinfo {volume} {634}},\ \bibinfo {pages} {69} (\bibinfo {year} {2024})}\BibitemShut {NoStop}%
\bibitem [{\citenamefont {Song}\ \emph {et~al.}(2024)\citenamefont {Song}, \citenamefont {Hao}, \citenamefont {Yan}, \citenamefont {Fang}, \citenamefont {Xu}, \citenamefont {Tong},\ and\ \citenamefont {Zhang}}]{song2024observation}%
  \BibitemOpen
  \bibfield  {author} {\bibinfo {author} {\bibfnamefont {G.}~\bibnamefont {Song}}, \bibinfo {author} {\bibfnamefont {H.}~\bibnamefont {Hao}}, \bibinfo {author} {\bibfnamefont {S.}~\bibnamefont {Yan}}, \bibinfo {author} {\bibfnamefont {S.}~\bibnamefont {Fang}}, \bibinfo {author} {\bibfnamefont {W.}~\bibnamefont {Xu}}, \bibinfo {author} {\bibfnamefont {L.}~\bibnamefont {Tong}},\ and\ \bibinfo {author} {\bibfnamefont {J.}~\bibnamefont {Zhang}},\ }\bibfield  {title} {\bibinfo {title} {{Observation of chirality transfer in twisted few-layer graphene}},\ }\href {https://doi.org/10.1021/acsnano.4c01934} {\bibfield  {journal} {\bibinfo  {journal} {ACS Nano}\ }\textbf {\bibinfo {volume} {18}},\ \bibinfo {pages} {17578} (\bibinfo {year} {2024})}\BibitemShut {NoStop}%
\bibitem [{\citenamefont {de~Almeida}\ \emph {et~al.}(2026)\citenamefont {de~Almeida}, \citenamefont {Kort-Kamp},\ and\ \citenamefont {Scheurer}}]{de2026high}%
  \BibitemOpen
  \bibfield  {author} {\bibinfo {author} {\bibfnamefont {J.~O.}\ \bibnamefont {de~Almeida}}, \bibinfo {author} {\bibfnamefont {W.~J.}\ \bibnamefont {Kort-Kamp}},\ and\ \bibinfo {author} {\bibfnamefont {M.~S.}\ \bibnamefont {Scheurer}},\ }\bibfield  {title} {\bibinfo {title} {{High-harmonic generation in systems with chiral Bloch states: application to rhombohedral graphene}},\ }\bibfield  {journal} {\bibinfo  {journal} {arXiv}\ }\href {https://doi.org/10.48550/arXiv.2604.11984} {10.48550/arXiv.2604.11984} (\bibinfo {year} {2026})\BibitemShut {NoStop}%
\bibitem [{\citenamefont {Khaliji}\ \emph {et~al.}(2022)\citenamefont {Khaliji}, \citenamefont {Martin-Moreno}, \citenamefont {Avouris}, \citenamefont {Oh},\ and\ \citenamefont {Low}}]{khaliji2022twisted}%
  \BibitemOpen
  \bibfield  {author} {\bibinfo {author} {\bibfnamefont {K.}~\bibnamefont {Khaliji}}, \bibinfo {author} {\bibfnamefont {L.}~\bibnamefont {Martin-Moreno}}, \bibinfo {author} {\bibfnamefont {P.}~\bibnamefont {Avouris}}, \bibinfo {author} {\bibfnamefont {S.-H.}\ \bibnamefont {Oh}},\ and\ \bibinfo {author} {\bibfnamefont {T.}~\bibnamefont {Low}},\ }\bibfield  {title} {\bibinfo {title} {{Twisted two-dimensional material stacks for polarization optics}},\ }\href {https://doi.org/10.1103/PhysRevLett.128.193902} {\bibfield  {journal} {\bibinfo  {journal} {Phys. Rev. Lett.}\ }\textbf {\bibinfo {volume} {128}},\ \bibinfo {pages} {193902} (\bibinfo {year} {2022})}\BibitemShut {NoStop}%
\bibitem [{\citenamefont {Tung}\ \emph {et~al.}(2017)\citenamefont {Tung}, \citenamefont {Chen}, \citenamefont {Jheng},\ and\ \citenamefont {Hung}}]{tung2017origin}%
  \BibitemOpen
  \bibfield  {author} {\bibinfo {author} {\bibfnamefont {H.~T.}\ \bibnamefont {Tung}}, \bibinfo {author} {\bibfnamefont {Y.~K.}\ \bibnamefont {Chen}}, \bibinfo {author} {\bibfnamefont {P.~L.}\ \bibnamefont {Jheng}},\ and\ \bibinfo {author} {\bibfnamefont {Y.~C.}\ \bibnamefont {Hung}},\ }\bibfield  {title} {\bibinfo {title} {{Origin and manipulation of band gaps in three-dimensional dielectric helix structures}},\ }\href {https://doi.org/10.1364/OE.25.017627} {\bibfield  {journal} {\bibinfo  {journal} {Opt. Express}\ }\textbf {\bibinfo {volume} {25}},\ \bibinfo {pages} {17627} (\bibinfo {year} {2017})}\BibitemShut {NoStop}%
\bibitem [{\citenamefont {Wu}\ \emph {et~al.}(2020)\citenamefont {Wu}, \citenamefont {Zhang},\ and\ \citenamefont {Das~Sarma}}]{wu2020three}%
  \BibitemOpen
  \bibfield  {author} {\bibinfo {author} {\bibfnamefont {F.}~\bibnamefont {Wu}}, \bibinfo {author} {\bibfnamefont {R.~X.}\ \bibnamefont {Zhang}},\ and\ \bibinfo {author} {\bibfnamefont {S.}~\bibnamefont {Das~Sarma}},\ }\bibfield  {title} {\bibinfo {title} {{Three-dimensional topological twistronics}},\ }\href {https://doi.org/10.1103/PhysRevResearch.2.022010} {\bibfield  {journal} {\bibinfo  {journal} {Phys. Rev. Res.}\ }\textbf {\bibinfo {volume} {2}},\ \bibinfo {pages} {022010} (\bibinfo {year} {2020})}\BibitemShut {NoStop}%
\bibitem [{\citenamefont {Wang}\ and\ \citenamefont {Huang}(2025)}]{wang2025decomposing}%
  \BibitemOpen
  \bibfield  {author} {\bibinfo {author} {\bibfnamefont {C.}~\bibnamefont {Wang}}\ and\ \bibinfo {author} {\bibfnamefont {H.}~\bibnamefont {Huang}},\ }\bibfield  {title} {\bibinfo {title} {{Decomposing electronic structures in twisted multilayers: Bridging spectra and incommensurate wave functions}},\ }\href {https://doi.org/10.1103/PhysRevB.111.195161} {\bibfield  {journal} {\bibinfo  {journal} {Phys. Rev. B}\ }\textbf {\bibinfo {volume} {111}},\ \bibinfo {pages} {195161} (\bibinfo {year} {2025})}\BibitemShut {NoStop}%
\bibitem [{\citenamefont {Phong}\ \emph {et~al.}(2025)\citenamefont {Phong}, \citenamefont {Kunkelmann}, \citenamefont {De~Beule}, \citenamefont {Al~Ezzi}, \citenamefont {Slager}, \citenamefont {Adam},\ and\ \citenamefont {Mele}}]{phong2025squeezing}%
  \BibitemOpen
  \bibfield  {author} {\bibinfo {author} {\bibfnamefont {V.~T.}\ \bibnamefont {Phong}}, \bibinfo {author} {\bibfnamefont {K.}~\bibnamefont {Kunkelmann}}, \bibinfo {author} {\bibfnamefont {C.}~\bibnamefont {De~Beule}}, \bibinfo {author} {\bibfnamefont {M.~M.}\ \bibnamefont {Al~Ezzi}}, \bibinfo {author} {\bibfnamefont {R.~J.}\ \bibnamefont {Slager}}, \bibinfo {author} {\bibfnamefont {S.}~\bibnamefont {Adam}},\ and\ \bibinfo {author} {\bibfnamefont {E.}~\bibnamefont {Mele}},\ }\bibfield  {title} {\bibinfo {title} {{Squeezing quantum states in three-dimensional twisted crystals}},\ }\href {https://doi.org/10.1103/cj2q-f9q2} {\bibfield  {journal} {\bibinfo  {journal} {Phys. Rev. B}\ }\textbf {\bibinfo {volume} {111}},\ \bibinfo {pages} {245156} (\bibinfo {year} {2025})}\BibitemShut {NoStop}%
\bibitem [{\citenamefont {Park}\ \emph {et~al.}(2026)\citenamefont {Park}, \citenamefont {Kim}, \citenamefont {Hwang},\ and\ \citenamefont {Min}}]{park2026magnetoplasmons}%
  \BibitemOpen
  \bibfield  {author} {\bibinfo {author} {\bibfnamefont {J.}~\bibnamefont {Park}}, \bibinfo {author} {\bibfnamefont {T.}~\bibnamefont {Kim}}, \bibinfo {author} {\bibfnamefont {E.}~\bibnamefont {Hwang}},\ and\ \bibinfo {author} {\bibfnamefont {H.}~\bibnamefont {Min}},\ }\bibfield  {title} {\bibinfo {title} {{Magnetoplasmons in N-layer structures}},\ }\bibfield  {journal} {\bibinfo  {journal} {arXiv}\ }\href {https://doi.org/10.48550/arXiv.2602.12722} {10.48550/arXiv.2602.12722} (\bibinfo {year} {2026})\BibitemShut {NoStop}%
\bibitem [{\citenamefont {Kazinski}\ and\ \citenamefont {Korolev}(2022)}]{kazinski2022scattering}%
  \BibitemOpen
  \bibfield  {author} {\bibinfo {author} {\bibfnamefont {P.~O.}\ \bibnamefont {Kazinski}}\ and\ \bibinfo {author} {\bibfnamefont {P.~S.}\ \bibnamefont {Korolev}},\ }\bibfield  {title} {\bibinfo {title} {{Scattering of plane-wave and twisted photons by helical media}},\ }\href {https://doi.org/10.1088/1751-8121/ac89ea} {\bibfield  {journal} {\bibinfo  {journal} {J. Phys. A: Math. Theor.}\ }\textbf {\bibinfo {volume} {55}},\ \bibinfo {pages} {395301} (\bibinfo {year} {2022})}\BibitemShut {NoStop}%
\bibitem [{\citenamefont {Tani}\ \emph {et~al.}(2023)\citenamefont {Tani}, \citenamefont {Kawakami},\ and\ \citenamefont {Koshino}}]{tani2023perpendicular}%
  \BibitemOpen
  \bibfield  {author} {\bibinfo {author} {\bibfnamefont {T.}~\bibnamefont {Tani}}, \bibinfo {author} {\bibfnamefont {T.}~\bibnamefont {Kawakami}},\ and\ \bibinfo {author} {\bibfnamefont {M.}~\bibnamefont {Koshino}},\ }\bibfield  {title} {\bibinfo {title} {{Perpendicular electronic transport and moire-induced resonance in twisted interfaces of three-dimensional graphite}},\ }\href {https://doi.org/10.1103/PhysRevB.108.165422} {\bibfield  {journal} {\bibinfo  {journal} {Phys. Rev. B}\ }\textbf {\bibinfo {volume} {108}},\ \bibinfo {pages} {165422} (\bibinfo {year} {2023})}\BibitemShut {NoStop}%
\bibitem [{\citenamefont {Crosse}\ and\ \citenamefont {Moon}(2021)}]{crosse2021faraday}%
  \BibitemOpen
  \bibfield  {author} {\bibinfo {author} {\bibfnamefont {J.}~\bibnamefont {Crosse}}\ and\ \bibinfo {author} {\bibfnamefont {P.}~\bibnamefont {Moon}},\ }\bibfield  {title} {\bibinfo {title} {{Faraday rotations, ellipticity, and circular dichroism in magneto-optical spectrum of moire superlattices}},\ }\href {https://doi.org/10.1088/1674-1056/ac051f} {\bibfield  {journal} {\bibinfo  {journal} {Chin. Phys. B}\ }\textbf {\bibinfo {volume} {30}},\ \bibinfo {pages} {077803} (\bibinfo {year} {2021})}\BibitemShut {NoStop}%
\bibitem [{\citenamefont {Mead}\ \emph {et~al.}(2025)\citenamefont {Mead}, \citenamefont {Talkington}, \citenamefont {Chen}, \citenamefont {Mallick}, \citenamefont {Chu}, \citenamefont {Han}, \citenamefont {Yang}, \citenamefont {Kim}, \citenamefont {Brahlek}, \citenamefont {Mele},\ and\ \citenamefont {Wu}}]{mead2025terahertz}%
  \BibitemOpen
  \bibfield  {author} {\bibinfo {author} {\bibfnamefont {B.~F.}\ \bibnamefont {Mead}}, \bibinfo {author} {\bibfnamefont {S.}~\bibnamefont {Talkington}}, \bibinfo {author} {\bibfnamefont {A.~H.}\ \bibnamefont {Chen}}, \bibinfo {author} {\bibfnamefont {D.}~\bibnamefont {Mallick}}, \bibinfo {author} {\bibfnamefont {Z.}~\bibnamefont {Chu}}, \bibinfo {author} {\bibfnamefont {X.}~\bibnamefont {Han}}, \bibinfo {author} {\bibfnamefont {S.~J.}\ \bibnamefont {Yang}}, \bibinfo {author} {\bibfnamefont {C.~J.}\ \bibnamefont {Kim}}, \bibinfo {author} {\bibfnamefont {M.}~\bibnamefont {Brahlek}}, \bibinfo {author} {\bibfnamefont {E.~J.}\ \bibnamefont {Mele}},\ and\ \bibinfo {author} {\bibfnamefont {L.}~\bibnamefont {Wu}},\ }\bibfield  {title} {\bibinfo {title} {{Terahertz Landau level spectroscopy of Dirac fermions in millimeter-scale twisted bilayer graphene}},\ }\href {https://doi.org/10.1103/5s4k-5jtj} {\bibfield  {journal} {\bibinfo  {journal} {Phys. Rev. B}\ }\textbf {\bibinfo {volume} {112}},\ \bibinfo {pages} {205116}
  (\bibinfo {year} {2025})}\BibitemShut {NoStop}%
\bibitem [{\citenamefont {Kelardeh}\ \emph {et~al.}(2014)\citenamefont {Kelardeh}, \citenamefont {Apalkov},\ and\ \citenamefont {Stockman}}]{kelardeh2014wannier}%
  \BibitemOpen
  \bibfield  {author} {\bibinfo {author} {\bibfnamefont {H.~K.}\ \bibnamefont {Kelardeh}}, \bibinfo {author} {\bibfnamefont {V.}~\bibnamefont {Apalkov}},\ and\ \bibinfo {author} {\bibfnamefont {M.~I.}\ \bibnamefont {Stockman}},\ }\bibfield  {title} {\bibinfo {title} {{Wannier-Stark states of graphene in strong electric field}},\ }\href {https://doi.org/10.1103/PhysRevB.90.085313} {\bibfield  {journal} {\bibinfo  {journal} {Phys. Rev. B}\ }\textbf {\bibinfo {volume} {90}},\ \bibinfo {pages} {085313} (\bibinfo {year} {2014})}\BibitemShut {NoStop}%
\bibitem [{\citenamefont {Iafrate}\ and\ \citenamefont {Sokolov}(2020)}]{iafrate2020bloch}%
  \BibitemOpen
  \bibfield  {author} {\bibinfo {author} {\bibfnamefont {G.}~\bibnamefont {Iafrate}}\ and\ \bibinfo {author} {\bibfnamefont {V.}~\bibnamefont {Sokolov}},\ }\bibfield  {title} {\bibinfo {title} {{The Bloch Electron Response to Electric Fields: Application to Graphene}},\ }\href {https://doi.org/10.1002/pssb.201900660} {\bibfield  {journal} {\bibinfo  {journal} {Phys. Status Solidi B}\ }\textbf {\bibinfo {volume} {257}},\ \bibinfo {pages} {1900660} (\bibinfo {year} {2020})}\BibitemShut {NoStop}%
\bibitem [{\citenamefont {De~Beule}\ \emph {et~al.}(2024)\citenamefont {De~Beule}, \citenamefont {Gassner}, \citenamefont {Talkington},\ and\ \citenamefont {Mele}}]{debeule2024floquet}%
  \BibitemOpen
  \bibfield  {author} {\bibinfo {author} {\bibfnamefont {C.}~\bibnamefont {De~Beule}}, \bibinfo {author} {\bibfnamefont {S.}~\bibnamefont {Gassner}}, \bibinfo {author} {\bibfnamefont {S.}~\bibnamefont {Talkington}},\ and\ \bibinfo {author} {\bibfnamefont {E.~J.}\ \bibnamefont {Mele}},\ }\bibfield  {title} {\bibinfo {title} {{Floquet-Bloch theory for nonperturbative response to a static drive}},\ }\href {https://doi.org/10.1103/PhysRevB.109.235421} {\bibfield  {journal} {\bibinfo  {journal} {Phys. Rev. B}\ }\textbf {\bibinfo {volume} {109}},\ \bibinfo {pages} {235421} (\bibinfo {year} {2024})}\BibitemShut {NoStop}%
\bibitem [{\citenamefont {Joy}\ \emph {et~al.}(2020)\citenamefont {Joy}, \citenamefont {Khalid},\ and\ \citenamefont {Skinner}}]{joy2020transparent}%
  \BibitemOpen
  \bibfield  {author} {\bibinfo {author} {\bibfnamefont {S.}~\bibnamefont {Joy}}, \bibinfo {author} {\bibfnamefont {S.}~\bibnamefont {Khalid}},\ and\ \bibinfo {author} {\bibfnamefont {B.}~\bibnamefont {Skinner}},\ }\bibfield  {title} {\bibinfo {title} {{Transparent mirror effect in twist-angle-disordered bilayer graphene}},\ }\href {https://doi.org/10.1103/PhysRevResearch.2.043416} {\bibfield  {journal} {\bibinfo  {journal} {Phys. Rev. Res.}\ }\textbf {\bibinfo {volume} {2}},\ \bibinfo {pages} {043416} (\bibinfo {year} {2020})}\BibitemShut {NoStop}%
\bibitem [{\citenamefont {Talkington}\ \emph {et~al.}(2026)\citenamefont {Talkington}, \citenamefont {Mallick}, \citenamefont {Chen}, \citenamefont {Mead}, \citenamefont {Yang}, \citenamefont {Kim}, \citenamefont {Adam}, \citenamefont {Wu}, \citenamefont {Brahlek},\ and\ \citenamefont {Mele}}]{talkington2025weak}%
  \BibitemOpen
  \bibfield  {author} {\bibinfo {author} {\bibfnamefont {S.}~\bibnamefont {Talkington}}, \bibinfo {author} {\bibfnamefont {D.}~\bibnamefont {Mallick}}, \bibinfo {author} {\bibfnamefont {A.~H.}\ \bibnamefont {Chen}}, \bibinfo {author} {\bibfnamefont {B.~F.}\ \bibnamefont {Mead}}, \bibinfo {author} {\bibfnamefont {S.~J.}\ \bibnamefont {Yang}}, \bibinfo {author} {\bibfnamefont {C.~J.}\ \bibnamefont {Kim}}, \bibinfo {author} {\bibfnamefont {S.}~\bibnamefont {Adam}}, \bibinfo {author} {\bibfnamefont {L.}~\bibnamefont {Wu}}, \bibinfo {author} {\bibfnamefont {M.}~\bibnamefont {Brahlek}},\ and\ \bibinfo {author} {\bibfnamefont {E.~J.}\ \bibnamefont {Mele}},\ }\bibfield  {title} {\bibinfo {title} {{Weak localization and universal conductance fluctuations in large-area twisted bilayer graphene}},\ }\href {https://doi.org/10.1103/xv5t-qvcm} {\bibfield  {journal} {\bibinfo  {journal} {Phys. Rev. B}\ }\textbf {\bibinfo {volume} {113}},\ \bibinfo {pages} {165430} (\bibinfo {year} {2026})}\BibitemShut {NoStop}%
\bibitem [{\citenamefont {Ollier}\ \emph {et~al.}(2023)\citenamefont {Ollier}, \citenamefont {Kisiel}, \citenamefont {Lu}, \citenamefont {Gysin}, \citenamefont {Poggio}, \citenamefont {Efetov},\ and\ \citenamefont {Meyer}}]{ollier2023energy}%
  \BibitemOpen
  \bibfield  {author} {\bibinfo {author} {\bibfnamefont {A.}~\bibnamefont {Ollier}}, \bibinfo {author} {\bibfnamefont {M.}~\bibnamefont {Kisiel}}, \bibinfo {author} {\bibfnamefont {X.}~\bibnamefont {Lu}}, \bibinfo {author} {\bibfnamefont {U.}~\bibnamefont {Gysin}}, \bibinfo {author} {\bibfnamefont {M.}~\bibnamefont {Poggio}}, \bibinfo {author} {\bibfnamefont {D.~K.}\ \bibnamefont {Efetov}},\ and\ \bibinfo {author} {\bibfnamefont {E.}~\bibnamefont {Meyer}},\ }\bibfield  {title} {\bibinfo {title} {{Energy dissipation on magic angle twisted bilayer graphene}},\ }\href {https://doi.org/10.1038/s42005-023-01441-4} {\bibfield  {journal} {\bibinfo  {journal} {Commun. Phys.}\ }\textbf {\bibinfo {volume} {6}},\ \bibinfo {pages} {344} (\bibinfo {year} {2023})}\BibitemShut {NoStop}%
\bibitem [{\citenamefont {Talkington}\ and\ \citenamefont {Claassen}(2024)}]{talkington2024linear}%
  \BibitemOpen
  \bibfield  {author} {\bibinfo {author} {\bibfnamefont {S.}~\bibnamefont {Talkington}}\ and\ \bibinfo {author} {\bibfnamefont {M.}~\bibnamefont {Claassen}},\ }\bibfield  {title} {\bibinfo {title} {{Linear and non-linear response of quadratic Lindbladians}},\ }\href {https://doi.org/10.1038/s41535-024-00709-4} {\bibfield  {journal} {\bibinfo  {journal} {npj Quantum Mater.}\ }\textbf {\bibinfo {volume} {9}},\ \bibinfo {pages} {104} (\bibinfo {year} {2024})}\BibitemShut {NoStop}%
\bibitem [{\citenamefont {Esparza}\ and\ \citenamefont {Juricic}(2025)}]{esparza2025exceptional}%
  \BibitemOpen
  \bibfield  {author} {\bibinfo {author} {\bibfnamefont {J.~P.}\ \bibnamefont {Esparza}}\ and\ \bibinfo {author} {\bibfnamefont {V.}~\bibnamefont {Juricic}},\ }\bibfield  {title} {\bibinfo {title} {{Exceptional magic angles in non-Hermitian twisted bilayer graphene}},\ }\href {https://doi.org/10.1103/dl59-vl7v} {\bibfield  {journal} {\bibinfo  {journal} {Phys. Rev. Lett.}\ }\textbf {\bibinfo {volume} {134}},\ \bibinfo {pages} {226602} (\bibinfo {year} {2025})}\BibitemShut {NoStop}%
\bibitem [{\citenamefont {Rhim}\ and\ \citenamefont {Yang}(2021)}]{rhim2021singular}%
  \BibitemOpen
  \bibfield  {author} {\bibinfo {author} {\bibfnamefont {J.~W.}\ \bibnamefont {Rhim}}\ and\ \bibinfo {author} {\bibfnamefont {B.~J.}\ \bibnamefont {Yang}},\ }\bibfield  {title} {\bibinfo {title} {{Singular flat bands}},\ }\href {https://doi.org/10.1080/23746149.2021.1901606} {\bibfield  {journal} {\bibinfo  {journal} {Adv. Phys. X}\ }\textbf {\bibinfo {volume} {6}},\ \bibinfo {pages} {1901606} (\bibinfo {year} {2021})}\BibitemShut {NoStop}%
\bibitem [{\citenamefont {Talkington}\ and\ \citenamefont {Claassen}(2022)}]{talkington2022dissipation}%
  \BibitemOpen
  \bibfield  {author} {\bibinfo {author} {\bibfnamefont {S.}~\bibnamefont {Talkington}}\ and\ \bibinfo {author} {\bibfnamefont {M.}~\bibnamefont {Claassen}},\ }\bibfield  {title} {\bibinfo {title} {{Dissipation-induced flat bands}},\ }\href {https://doi.org/10.1103/PhysRevB.106.L161109} {\bibfield  {journal} {\bibinfo  {journal} {Phys. Rev. B}\ }\textbf {\bibinfo {volume} {106}},\ \bibinfo {pages} {L161109} (\bibinfo {year} {2022})}\BibitemShut {NoStop}%
\bibitem [{\citenamefont {Regnault}\ \emph {et~al.}(2022)\citenamefont {Regnault}, \citenamefont {Xu}, \citenamefont {Li}, \citenamefont {Ma}, \citenamefont {Jovanovic}, \citenamefont {Yazdani}, \citenamefont {Parkin}, \citenamefont {Felser}, \citenamefont {Schoop}, \citenamefont {Ong} \emph {et~al.}}]{regnault2022catalogue}%
  \BibitemOpen
  \bibfield  {author} {\bibinfo {author} {\bibfnamefont {N.}~\bibnamefont {Regnault}}, \bibinfo {author} {\bibfnamefont {Y.}~\bibnamefont {Xu}}, \bibinfo {author} {\bibfnamefont {M.~R.}\ \bibnamefont {Li}}, \bibinfo {author} {\bibfnamefont {D.~S.}\ \bibnamefont {Ma}}, \bibinfo {author} {\bibfnamefont {M.}~\bibnamefont {Jovanovic}}, \bibinfo {author} {\bibfnamefont {A.}~\bibnamefont {Yazdani}}, \bibinfo {author} {\bibfnamefont {S.~S.~P.}\ \bibnamefont {Parkin}}, \bibinfo {author} {\bibfnamefont {C.}~\bibnamefont {Felser}}, \bibinfo {author} {\bibfnamefont {L.~M.}\ \bibnamefont {Schoop}}, \bibinfo {author} {\bibfnamefont {N.~P.}\ \bibnamefont {Ong}}, \emph {et~al.},\ }\bibfield  {title} {\bibinfo {title} {{Catalogue of flat-band stoichiometric materials}},\ }\href {https://doi.org/10.1038/s41586-022-04519-1} {\bibfield  {journal} {\bibinfo  {journal} {Nature}\ }\textbf {\bibinfo {volume} {603}},\ \bibinfo {pages} {824} (\bibinfo {year} {2022})}\BibitemShut {NoStop}%
\bibitem [{\citenamefont {Margetis}\ \emph {et~al.}(2024)\citenamefont {Margetis}, \citenamefont {Gomez-Santos},\ and\ \citenamefont {Stauber}}]{margetis2024optical}%
  \BibitemOpen
  \bibfield  {author} {\bibinfo {author} {\bibfnamefont {D.}~\bibnamefont {Margetis}}, \bibinfo {author} {\bibfnamefont {G.}~\bibnamefont {Gomez-Santos}},\ and\ \bibinfo {author} {\bibfnamefont {T.}~\bibnamefont {Stauber}},\ }\bibfield  {title} {\bibinfo {title} {{Optical response of alternating twisted trilayer graphene}},\ }\href {https://doi.org/10.1103/PhysRevB.110.205144} {\bibfield  {journal} {\bibinfo  {journal} {Phys. Rev. B}\ }\textbf {\bibinfo {volume} {110}},\ \bibinfo {pages} {205144} (\bibinfo {year} {2024})}\BibitemShut {NoStop}%
\bibitem [{\citenamefont {Xiao}\ \emph {et~al.}(2026{\natexlab{b}})\citenamefont {Xiao}, \citenamefont {Xiao}, \citenamefont {Zhai},\ and\ \citenamefont {Yao}}]{xiao2026interlayer}%
  \BibitemOpen
  \bibfield  {author} {\bibinfo {author} {\bibfnamefont {C.}~\bibnamefont {Xiao}}, \bibinfo {author} {\bibfnamefont {C.}~\bibnamefont {Xiao}}, \bibinfo {author} {\bibfnamefont {D.}~\bibnamefont {Zhai}},\ and\ \bibinfo {author} {\bibfnamefont {W.}~\bibnamefont {Yao}},\ }\bibfield  {title} {\bibinfo {title} {{Interlayer electric multipole Hall effect in twisted multilayers}},\ }\bibfield  {journal} {\bibinfo  {journal} {arXiv}\ }\href {https://doi.org/10.48550/arXiv.2606.28205} {10.48550/arXiv.2606.28205} (\bibinfo {year} {2026}{\natexlab{b}})\BibitemShut {NoStop}%
\bibitem [{\citenamefont {Yokoshi}\ and\ \citenamefont {Kato}(2026)}]{yokoshi2026optical}%
  \BibitemOpen
  \bibfield  {author} {\bibinfo {author} {\bibfnamefont {N.}~\bibnamefont {Yokoshi}}\ and\ \bibinfo {author} {\bibfnamefont {A.}~\bibnamefont {Kato}},\ }\bibfield  {title} {\bibinfo {title} {{Optical vortex probe of loop-current chirality in moire materials}},\ }\href {https://doi.org/10.1103/dlgp-bgbb} {\bibfield  {journal} {\bibinfo  {journal} {Phys. Rev. B}\ }\textbf {\bibinfo {volume} {113}},\ \bibinfo {pages} {245303} (\bibinfo {year} {2026})}\BibitemShut {NoStop}%
\end{thebibliography}
%

\end{document}